\documentclass[11pt]{article}
\usepackage{amssymb}
\usepackage{mathtools}
\usepackage{amsmath}
\usepackage{amstext}
\usepackage{graphicx,epsfig}
\usepackage{epsfig}
\usepackage{verbatim} 	
\usepackage{caption}
\usepackage{fancybox}
\usepackage{slashed}
\usepackage{jheppub}
\usepackage{color}
\usepackage{ulem}
\usepackage{enumitem}
\usepackage{subfigure}
\usepackage{parskip}
\usepackage{dsfont}
\usepackage{tabu}
\usepackage{tikz}
\usepackage{booktabs}
\usepackage[numbers,sort&compress]{natbib}
\usepackage{cancel}

\usepackage[font={small,it}]{caption}

\newcommand{\Comment}[1]{{}}
\definecolor{darkblue}{rgb}{0.15,0.35,0.55}
\definecolor{reddish}{rgb}{0.65, 0.2, 0.2}
\definecolor{taiga}{cmyk}{0.68,0,0.56,0.49}
\usepackage[linktocpage=true]{hyperref}
\hypersetup{
colorlinks=true,
citecolor=darkblue,
linkcolor=reddish,
urlcolor=taiga,
pdfauthor={},
pdftitle={},
pdfsubject={}
}

\newcommand{\D}{{\rm d}}

\newcommand{\MP}{M_{\rm Pl}}

\newcommand{\bfk}{{\bf k}}

\newcommand{\bfx}{{\bf x}}

\newcommand{\bfnabla}{\nabla}

\newcommand{\be}{\begin{eqnarray} }
\newcommand{\ee}{\end{eqnarray} }
\newcommand{\bs}{\begin{split} }
\newcommand{\es}{\end{split} }

\newcommand{\Mpl}{M_{\mathrm{Pl}}}

\newcommand{\e}{{\rm e}}

\def\eq#1{{(\ref{#1})}}

\title{}
\author{}

\numberwithin{equation}{section}

\begin{document}
%
\renewcommand{\thefootnote}{\fnsymbol{footnote}}
~
\thispagestyle{empty}
\vspace{-2cm} {\flushright IFT-UAM/CSIC-21-93} \\
\vspace{2cm}
\begin{center}
{\huge\bf{
Large power spectrum and primordial black holes in the effective theory of inflation\\[0.3cm]}}
\end{center} 

\vspace{1.2truecm}

\begin{center}
{\fontsize{13.5}{18}\selectfont
Guillermo Ballesteros,${}^{\rm a,b}$ Sebasti\'an C\'espedes,${}^{\rm a,b}$ Luca Santoni${}^{\rm c}$
}
\end{center}
\vspace{.5truecm}

 \centerline{{\it ${}^{\rm a}$Instituto de F\'isica Te\'orica UAM/CSIC,}}
 \centerline{{\it Calle Nicol\'as Cabrera 13-15, Cantoblanco E-28049 Madrid, Spain}}
 
  \vspace{.3cm}
  
   \centerline{{\it ${}^{\rm b}$Departamento de F\'isica Te\'orica, Universidad Aut\'onoma de Madrid (UAM)}}
 \centerline{{\it Campus de Cantoblanco, E-28049 Madrid, Spain}}
 
  \vspace{.3cm}
  
  \centerline{{\it ${}^{\rm c}$Center for Theoretical Physics, Department of Physics,}}
 \centerline{{\it Columbia University, New York, NY 10027}}

\vspace{1.5cm}
\begin{abstract}
\noindent
We study the  generation of a large power spectrum, necessary for primordial black hole formation, within the effective theory of  single-field inflation. The mechanisms we consider include a transition into a ghost-inflation-like phase and scenarios where an exponentially growing mode is temporarily turned on.
In the cases we discuss, the enhancement in the power spectrum results from either a swift change in some effective coupling or a modification of the dispersion relation for the  perturbations, while  the background evolution remains unchanged and approximately de Sitter throughout  inflation.
The robustness of the results is guaranteed thanks to a weakly broken galileon  symmetry, which protects the effective couplings against large quantum corrections.  
We discuss how the enhancement of the power spectrum is related to the energy scale of the operators with weakly broken galileon invariance,
and study the limits imposed by strong coupling and the  validity of the perturbative expansion.
\end{abstract}

\begin{center} 

\vfill\flushleft
\noindent\rule{6cm}{0.4pt}\\
{\small  \tt guillermo.ballesteros@uam.es, sebastian.cespedes@uam.es, luca.santoni@columbia.edu}

\end{center}

\newpage

\tableofcontents

\renewcommand*{\thefootnote}{\arabic{footnote}}
\setcounter{footnote}{0}

\newpage

\section{Introduction}

The idea that primordial black holes (PBHs) could constitute all the dark matter of the Universe or part of it has gained a remarkable momentum in recent years. The most popular scenario for the production of black holes in the primordial Universe consists in the collapse of radiation overdensities,  which originate  from large quantum fluctuations seeded during inflation, see e.g.~\cite{Carr:1993aq,Ivanov:1994pa}. Several models devised with the aim of enhancing the primordial spectrum have been recently explored. Most of them have  a single inflationary field whose dynamics deviates from standard slow-roll, leading to a primordial spectrum peaking with a large value at some scale of (comoving) distance. See \cite{Ivanov:1994pa,Garcia-Bellido:2017mdw,Ballesteros:2017fsr, Cicoli:2018asa, Ozsoy:2018flq,  Mishra:2019pzq, Ballesteros:2019hus, Ballesteros:2020qam} for a representative sample of a popular class of models.

From a broader perspective, one can identify two physically distinct options that implement dynamical characteristics leading to large primordial fluctuations. The first possibility is a substantial change in the homogeneous background evolution of the Universe during inflation. Many concrete models that fall in this category---and in particular the ones mentioned above---feature a significant reduction of the slow-roll parameter $\varepsilon\equiv-\dot{H}/H^2$. Rather generically, this can lead to a breakdown of the slow-roll evolution and, possibly, a transition into an ultra slow-roll phase \cite{Germani:2017bcs,Motohashi:2017kbs,Ballesteros:2017fsr}. This kind of background dynamics induces a large effect in the small quantum fluctuations that are inherently associated to the background, generating a large primordial spectrum. A second possibility consists instead in some modification of the dynamics of the perturbations alone, while keeping the background evolution in a, rather featureless, inflationary evolution, see \cite{Ballesteros:2018wlw} (and also \cite{Palma:2020ejf,Fumagalli:2020adf}). In the present work, we will focus on this latter case, assuming a standard quasi-de Sitter spacetime during inflation, with approximately constant slow-roll parameters throughout.

For our analysis to be as general and model-independent as possible, we will work within the framework of the effective field theory (EFT) of single-field inflation \cite{Creminelli:2006xe,Cheung:2007st}. In addition, the robustness of our results  will be guaranteed by an approximate symmetry,  which protects the effective couplings
against potentially large quantum corrections.  This is in contrast with the more commonly explored models, mentioned earlier, in which a large power spectrum comes from a very small $\varepsilon$. In such models, the smallness of $\varepsilon$ generically  arises as a consequence of a fine tuning of the parameters in the (covariant) Lagrangian of the inflaton. This tuning can be a cause of concern because  it is in general not protected by any symmetry. 

Besides  our phenomenological motivation of PBHs as dark matter, the question of determining the largest primordial spectrum consistently allowed in the EFT of inflation {from different operators} is interesting on its own. Although a very large spectrum is physically irrelevant at large enough distance scales, due to stringent Cosmic Microwave Background (CMB) and Large Scale Structure (LSS) bounds, the size of the spectrum at smaller scales can be orders of magnitude larger. Indeed, even though there exist indirect, model-dependent and scale-dependent bounds (coming mainly from the absence of compelling evidence for the existence of PBHs), there are comoving scales where the primordial spectrum is essentially unconstrained. In particular, there are no bounds in the range $10^{12}$ Mpc -- $10^{14}$ Mpc, corresponding to PBHs that would have formed during radiation domination with masses between $10^{-12}\, M_\odot$ and $10^{-16}\, M_\odot$.\footnote{See \cite{Green:2020jor,Carr:2020gox} for recent compilations of PBH bounds.}

In models that do not rely on a reduction of $\varepsilon$, a large primordial spectrum is usually accompanied by  sizable interactions between fluctuations and significant non-Gaussianities.  This means that the primordial spectrum cannot be arbitrarily large without stepping over the regime of validity of the EFT. For instance, the partial-wave unitarity cutoff for models with an inflaton $\phi$ featuring a Lagrangian of the form $\int \D^4x \sqrt{-g}\,[\MP^2 R/2+ P(\phi, (\partial_\mu\phi)^2)]$ \cite{Armendariz-Picon:1999hyi} and a small speed of sound $c_s \ll 1$ is $\Lambda_\star \sim \sqrt{\MP H}\, \left(\varepsilon c_s^5\right)^{1/4}$. The  consistency requirement $H \ll \Lambda_\star$ can then be rephrased as  $c_s^4 \gg {\Delta_ \mathcal{\zeta}^2}$ \cite{Cheung:2007st}, setting a bound on the maximum primordial spectrum $\Delta_ \mathcal{\zeta}^2$ that can be achieved for a given (small) value of $c_s$. Since in the simplest, Gaussian, approximation the PBH abundance depends (exponentially) on the primordial spectrum, the relation  $c_s^4 \gg {\Delta_ \mathcal{\zeta}^2}$ may be used to constrain mechanisms of PBH formation in scenarios in which $c_s$ is small \cite{Ballesteros:2018wlw}. Assuming that a large $\Delta_\zeta^2$ is needed to generate an abundance of PBHs compatible with the dark matter density, it was argued in \cite{Ballesteros:2018wlw} that mechanisms based on a small $c_s$ in the realm of the vanilla EFT of inflation---i.e.\ with the kind of covariant action mentioned above---point towards a very small $\varepsilon$ and, possibly $\varepsilon \ll c_s^2$. This conclusion assumes small variations of the EFT coefficients during inflation, but it is nevertheless indicative of the difficulty of obtaining a large primordial spectrum in the EFT of inflation from a variation of $c_s$; a difficulty that becomes more severe as the desired spectrum is increased.

A potential way of circumventing this issue, which was already suggested in \cite{Ballesteros:2018wlw}, consists in invoking a modified dispersion relation for the primordial fluctuations, as it happens e.g.\ in ghost inflation \cite{ArkaniHamed:2003uz}. This changes the cutoff of the EFT through the appearance of a new scale, 
which is associated to the higher-dimensional terms of the action responsible for the modified dispersion relation. In the present paper we will explore this idea in more depth. One of  main results of our paper is the following:  we  find that a transition into a phase of ghost condensation results in an enhanced power spectrum, while  it raises the unitarity cutoff of the theory, which avoids the strong coupling issues arising from a small speed of sound and keeps  the theory for the perturbations weakly coupled during the evolution. 
The enhancement of the primordial spectrum in this case is in the form of a power law and scales generically as $(\Lambda_\star/H)^{3/2}$, where $\Lambda_\star$ represents the new raised cutoff.

To put the following discussion into context, we need to understand what we mean by a `large power spectrum'. As we mentioned before, the PBH abundance is exponentially sensitive to the primordial power spectrum, in the Gaussian approximation. It turns out that, with this approximation, ${\Delta_ \mathcal{\zeta}^2}\sim 10^{-2}$ is required in order for PBHs to account for all the dark matter. This is a crude approximation as it ignores several effects that may be important, depending on the model, notably: non-Gaussianities, stochastic dynamics, shape of the primordial spectrum, threshold for the formation of PBHs and equation of state of the Universe at the time of formation. For this reason, we will not focus our analysis on obtaining a specific value of ${\Delta_ \mathcal{\zeta}^2}$, but rather on primordial spectra that are orders of magnitude larger than the CMB one, in a broad range. Nevertheless, we can still use ${\Delta_ \mathcal{\zeta}^2}\sim 10^{-2}$ as a convenient benchmark, keeping in mind that smaller values of ${\Delta_ \mathcal{\zeta}^2}$---possibly orders of magnitude smaller---may be enough to account for all dark matter. In any case, ${\Delta_ \mathcal{\zeta}^2}\sim 10^{-2}$ (or any meaningful large value relevant in this context) is indeed very large in comparison to the primordial spectrum inferred from the CMB, which is about seven orders of magnitude smaller.

In order to explore  the possibilities for obtaining a large primordial spectrum in the EFT of inflation away from a reduction in $\varepsilon$,\footnote{Often implying a large second slow-roll parameter $\eta$  in concrete models, see e.g.\ \cite{Ballesteros:2017fsr}.} we will focus on  the operators $(\delta g^ {00})^2$ and $\delta K \delta g^{00}$ in the unitary gauge action for the perturbations
\cite{Creminelli:2006xe,Cheung:2007st}. Despite being higher order in derivatives, $\delta K \delta g^{00}$ can become as large as $(\delta g^ {00})^2$ on the inflationary background thanks to a weakly broken galileon (WBG) symmetry, in a way that  is stable under loop corrections \cite{Pirtskhalava:2015nla,Santoni:2018rrx}. 
The combination of these operators can make the sound speed become  very small and the dispersion relation for the perturbations, which in general reads
\begin{align}
\omega^2 = c_s^2 k^2+ \alpha \frac{k ^4}{a^2H^2} \,  ,
\label{drintro}
\end{align}
  dominated by the $k ^4$ term (see Eq.~\eqref{cs} for the expressions of $c_s$ and $\alpha$ in terms of the effective coefficients).

In addition to the transition into a ghost-inflation-like phase, we discuss another mechanism, which allows to increase more efficiently the power spectrum. This consists in pushing the sound speed beyond $c_s^2 = 0$, allowing for a transient phase with $c_s^2 < 0$.    In this case, the primordial spectrum grows exponentially (at specific comoving scales).

As we will discuss,  there is in principle nothing wrong with this temporary instability provided that the dynamics of modes with the largest physical momenta is controlled  by the $ k^4$ term. However, the validity of perturbation theory will restrict the duration of the phase with $c_s^2 < 0$ as well as  the amplitude of the change in $c_s^2$.  Phases with this type of instability have been discussed before see e.g. \cite{Creminelli:2006xe}, and more recently in \cite{Garcia-Saenz:2018ifx, Garcia-Saenz:2018vqf, Fumagalli:2019noh,Bjorkmo:2019qno,Ferreira:2020qkf,Palma:2020ejf,Fumagalli:2020adf,Fumagalli:2020nvq}.

The  paper is organized as follows. In Section \ref{IFTintro} we review the EFT of the (scalar) fluctuations of a single scalar field coupled to gravity---during inflation and at quadratic order in the fluctuations---including the general next-to-leading order term in spatial derivatives, see Eq.\ \eq{genact}.
In Section \ref{enhanced} we discuss, using the EFT, to what extent it is possible to enhance the power spectrum by means of a negative friction coefficient in the linearized equation for the perturbations.
In Section \ref{sec:srtogis} we discuss how a transition from a standard phase with linear dispersion relation for the perturbations, $\omega = c_s k$, into a phase with  quadratic  dispersion relation, $\omega = \sqrt{\alpha} \, k^2/(aH)$, can result  in an enhanced  primordial spectrum.
In particular, we compare this enhancement to the one that can be obtained (within the EFT) from a swift change in the slow-roll parameters assuming $\omega = c_s k$.

In Section~\ref{ima} we entertain the possibility that $c_s^2$ becomes temporarily negative during inflation, leading to an exponential enhancement of the perturbations. In Section \ref{sec:ng-pt} we discuss the relation between non-Gaussianities and the validity of the EFT.  In Section \ref{sec:conclude} we present our conclusions. 
Appendices~\ref{app:sizeoperators} and \ref{app:ng-pt} deepen on various aspects of the main text, while  we compare in Appendix~\ref{app:2-fields} our findings to previous results in the context of multi-field models.

\paragraph{Conventions:} We work in mostly-plus signature for the metric, $(-,+,+,+)$. The Hubble and slow-roll parameters are defined in cosmological time by $H\equiv {\dot{a}}/{a}$, $\varepsilon=-{\dot{H}}/{H^2}$ and $\eta={\dot{\varepsilon}}/({H\varepsilon})$.  Throughout the paper we will declare that slow-roll is satisfied if $\varepsilon$, $\eta$ and higher order slow-roll parameters are $\ll 1$.  The conformal time  is denoted with $\tau$ and is defined by $a\, \D \tau \equiv {\D t}$. Comoving spatial momenta living in Fourier space are denoted in bold font as $ \bfk$, and we define $k\equiv \vert\bfk\vert$. The reduced Planck mass is $\MP= 1/\sqrt{8\pi G}$.

\section{Preliminary effective field theory considerations} \label{IFTintro}

Let us start from the  EFT of a single scalar degree of freedom coupled to gravity in a FLRW spacetime, with background metric $\bar{g}_{\mu\nu}=\text{diag}(-1,a(t)^2,a(t)^2,a(t)^2)$, in unitary gauge \cite{Creminelli:2006xe,Cheung:2007st} (in which all the fluctuations are on the spacetime metric). 
Up to quadratic order in perturbations, we will focus on the following subset of operators:
\begin{equation}
\begin{split}
S = \int \D^4 x \sqrt{-g}  \bigg[
&	 \frac{\MP^2}{2} R - \MP^2(3H^2+\dot H) + \MP^2\dot H  g^{00} 
\\
&	+ \frac{1}{2}M_2^4 (\delta g^{00})^2
- \frac{1}{2}\hat{M}^3_1 \delta g^{00}\delta K   	 
  - \frac{1}{2}\bar{M}^2_2(\delta K)^2 
+ \ldots
\bigg] \, ,
\end{split}
\label{eftofi}
\end{equation}
where $K={K^\mu}_\mu$ denotes the trace of the extrinsic curvature $K_{\mu\nu}$ associated with the equal-time hypersurfaces of the spacetime foliation defined by the unitary gauge choice  \cite{Cheung:2007st}. The first line in \eqref{eftofi} is unambiguously fixed by the background dynamics. The second line contains some of the quadratic operators that contribute to the action up to  second order in the derivative expansion. The coefficients of these operators, such as $M_2^4$, $\hat{M}^3_1$ and $\bar{M}^2_2$ are actually arbitrary functions of time, due to the breaking of time diffeomorphisms.  
In the absence of any other symmetries, higher derivative operators,\footnote{By higher derivative operators we mean operators that in the covariant Lagrangian have more than one derivative per field (see Appendix \ref{app:sizeoperators}), or equivalently that in the unitary gauge language of \eqref{eftofi} have at least one derivative acting on the metric perturbations.  }  like $\delta g^{00}\delta K$ and $(\delta K)^2$, are usually subleading and the dynamics of the perturbations at quadratic order is dominated  by  $(\delta g^{00})^2$. In this paper, we are interested instead in situations where  $M_2^4(\delta g^{00})^2$ and  $\hat{M}^3_1\delta g^{00}\delta K$ are  of the same order on the FLRW background. At first sight, this seems in contrast with the spirit of the EFT. However, as shown in \cite{Pirtskhalava:2015nla,Santoni:2018rrx}, having $M_2^4 \sim  \hat{M}^3_1H $ (and hence both operators of the same order) is  possible in theories characterized by a WBG invariance.  As a bonus, a non-renormalization theorem, stemming from the weakly broken symmetry, guarantees that quantum corrections to $M_2^4 $ and $  \hat{M}^3_1 $ are parametrically suppressed.  For further details on models featuring a WBG symmetry, see Appendix \ref{app:sizeoperators}.\footnote{In principle, there are other operators in \eqref{eftofi} that we did not write that belong to the same class of operators with WBG symmetry. However, thanks to the non-renormalization theorem,
it is completely safe to set their tree-level couplings to `zero'   in \eqref{eftofi}.}
The operator $(\delta K)^2$ is instead of a different type. It is not protected by symmetries and it is always subleading compared, e.g.\ to $(\delta g^{00})^2$ in the Lagrangian.\footnote{Unless it enters through the combination $(\delta K)^2- \delta K^{\mu\nu}\delta K_{\mu\nu}$, which also belongs to the class of Horndeski operators with WBG symmetry \cite{Pirtskhalava:2015nla}.}
Nevertheless, it may end up providing the leading correction to the dispersion relation for the perturbations \cite{ArkaniHamed:2003uy,ArkaniHamed:2003uz,Baumann:2011su}.\footnote{As we will see, this can happen as a result  of cancellations between different coefficients in the dispersion relation. However, this is not a fine tuning thanks to the WBG symmetry.}  Since we will   discuss this option in the following, we have written explicitly this operator in \eqref{eftofi}. Note however that  there are  other operators of the same type that enter at the same order in derivatives and that, in principle, we should have written in \eqref{eftofi}. In fact, these operators are generated via loop corrections from the ones in \eqref{eftofi}. We have omitted them for simplicity because they do not change our results qualitatively.	
In this sense, $(\delta K)^2$ should be considered as a representative of this class  of operators. 
To recap, we will focus below on the action \eqref{eftofi} where the effective couplings  satisfy the following hierarchy,\footnote{Notice that $H$ is the only relevant scale for the size of (massless) metric fluctuations during inflation. Therefore, since $\delta K$ has dimension of mass 1, its expected order of magnitude is $\delta K\sim H$ on purely dimensional grounds.}
\begin{equation}
M_2^4 \sim  \hat{M}^3_1H \gg \bar{M}^2_2H^2 \, ,
\end{equation}
which is stable against quantum corrections.

To study the dynamics of the scalar  perturbations it is convenient to use the $\zeta$-gauge, defined by \cite{Maldacena:2002vr}:
\begin{equation}
\delta g_{ij} = a^2 \e^{2\zeta} \delta_{ij}\,.
\end{equation} 
After integrating out the non-dynamical components of the metric, $\delta g^{00}$ and $\delta g^{0i}$, from \eqref{eftofi}, one finds the following quadratic  action for $\zeta$,
\begin{equation} \label{genact}
S^{(2)}_\zeta = \int \D^4 x \,A\, a^3 \left[\dot \zeta^2 - c_s^2 \frac{(\bfnabla\zeta)^2}{a^2}  - \alpha\, \frac{(\bfnabla^2\zeta)^2}{H^2a^4}\right] \, ,
\end{equation}
where
\begin{align}
A & = \frac{\MP^2 \left(2 \MP^2 \dot H \left(2 \MP^2+3 \bar{M}_2^2\right)-4 M_2^4 \left(2 \MP^2+3 \bar{M}_2^2\right)-3 \hat{M}_1^6\right)}{2 \MP^2 \bar{M}_2^2 \dot H- 2\MP^2H^2 \left(2\MP^2 +3 \bar{M}_2^2 \right)+4 \MP^2 H \hat{M}_1^3-4 M_2^4 \bar{M}_2^2 - \hat{M}_1^6} \, ,
\label{coeffA}
\\
 c_s^2 & = \frac{1}{A}\left( - \MP^2 + \frac{\dot Y}{a}  \right) \,,
\\
 \alpha & = - \frac{2 \MP^2H^2 \bar{M}_2^2}{2 \MP^2 \dot H \left(2 \MP^2+3 \bar{M}_2^2\right)-4 M_2^4 \left(2 \MP^2+3 \bar{M}_2^2\right)-3 \hat{M}_1^6} \, ,
 \label{alphagen}
\end{align}
being
\begin{equation}
Y = -\frac{2 \MP^4 a \left(H\left(2 \MP^2+3 \bar{M}_2^2\right)-\hat{M}_1^3\right)}{2 \MP^2 \bar{M}_2^2 \dot H- 2\MP^2H^2 \left(2\MP^2 +3 \bar{M}_2^2 \right)+4 \MP^2 H \hat{M}_1^3-4 M_2^4 \bar{M}_2^2 - \hat{M}_1^6} \, .
\label{coeffY}
\end{equation}
The possible values for the coefficients are restricted by the validity of the effective description  (see, e.g., \cite{Pirtskhalava:2015nla,Santoni:2018rrx} and Appendix~\ref{app:sizeoperators} below):
\begin{equation}
\label{cond1}
\varepsilon\MP^2 H^2 \lesssim M_2^4 , \,   \hat{M}^3_1H  \lesssim \MP^2 H^2 \, ,
\qquad
\varepsilon^{2/3}\MP^{4/3} H^{2/3} \lesssim \bar{M}^2_2     \lesssim \MP^{4/3} H^{2/3} \, ,
\end{equation}
where $\varepsilon\equiv -\dot{H}/H^2\ll1$.
To make our expressions simpler, we will work  in the decoupling limit (i.e.~the regime  where   the scalar mode is decoupled from the metric perturbations \cite{Cheung:2007st}),\footnote{To explore the whole range of values for the parameters in \eqref{cond1},  one would need to  take into account the coupling to the metric perturbations. However, this would not change  qualitatively our conclusions, but it would complicate unnecessarily the expressions for $c_s$, $\alpha$, etc..} which applies if \cite{Pirtskhalava:2015nla} 
\begin{equation}
 M_2^4 , \,   \hat{M}^3_1H  \ll \MP^2 H^2 \, ,
\qquad
 \bar{M}^2_2     \ll \MP^{4/3} H^{2/3} \, .
\label{M2Mb2h00}
\end{equation}
 Using \eqref{M2Mb2h00}, Eqs.~\eqref{coeffA}--\eqref{coeffY} simplify considerably and the quadratic action for $\zeta$  reduces to
\begin{equation}
S^{(2)}_\zeta = \int \D^4 x \, a^3 H^{-2} \left(2 M_2^4 - \MP^2\dot H \right)  \left[  \dot \zeta^2 - c_s^2 \frac{(\bfnabla\zeta)^2}{a^2}  - \alpha \frac{(\bfnabla^2\zeta)^2}{H^2a^4}\right] \, ,
\label{ghostL2}
\end{equation}
where now
\begin{equation}
c_s^2 = \frac{-2\MP^2 \dot H  +  \hat{M}_1^3 H  + \partial_t (\hat{M}_1^3) }{ 2( 2 M_2^4 - \MP^2\dot H )} \, ,
\qquad
\alpha = \frac{\bar{M}_2^2H^2}{2(2 M_2^4 - \MP^2\dot H )} \, .
\label{cs}
\end{equation}
Note that, even if the decoupling limit applies in the whole range \eqref{M2Mb2h00}, in the following we will mainly assume
\begin{equation}
M_2^4 \sim  \hat{M}^3_1H  \sim \varepsilon \MP^2H^2  \, ,
\qquad\qquad
\bar{M}_2^2  \sim \varepsilon^{2/3} \MP^{4/3}H^{2/3}  \, ,
\label{M2Mb2h}
\end{equation}
which corresponds to cases where the inflationary background dynamics is mostly driven by the scalar's potential \cite{Pirtskhalava:2015nla,Pirtskhalava:2015zwa}. If this latter hierarchy is satisfied, then\footnote{We stress again  that the effective theory \eqref{eftofi} will contain in general  other operators of the same type of  $(\delta K)^2$ (even if we did not write them explicitly in \eqref{eftofi})  that  contribute to $(\bfnabla^2\zeta)^2$ in \eqref{ghostL2} (and enter at the same scale of $(\delta K)^2$). These will change the explicit expression of  $\alpha$ in \eqref{cs}, but will not change its dependence on the relevant scales of the problem, Eq.~\eqref{alphaL3}, which is what we will be using in the following.  }
\begin{equation}
\alpha \sim \left(H / \Lambda_3 \right)^2 \,,
\label{alphaL3}
\end{equation} 
where we defined the scale $\Lambda_3$  as
\begin{equation} \label{Lambda3}
\Lambda_3\sim  \varepsilon^{1/6} (\MP H^2)^{1/3} \,.
\end{equation} 
In theories with a  WBG invariance,  $\Lambda_3$ is precisely the scale that suppresses the higher derivative operators; see, e.g., \cite{Pirtskhalava:2015nla,Pirtskhalava:2015zwa} and Appendix~\ref{app:sizeoperators} below.

Note that in \eqref{ghostL2} the quadratic operator $(\bfnabla^2\zeta)^2$ is the first of a series of terms of the form $(\bfnabla^{n}\zeta)^2$. We did not write explicitly these operators for $n>2$ because, as required by the consistency of the derivative expansion, they are increasingly subleading at low (comoving) momenta as $n$ grows, provided that
\begin{equation}
k \tau \lesssim \frac{1}{\sqrt{\alpha}} \, ,
\label{derexp}
\end{equation}
where $\tau$ is the conformal time and where we used the slow-roll approximation to write $\tau\sim(aH)^{-1}$.
Thus, at any given $\tau$, only modes with comoving momentum $k$ satisfying \eqref{derexp} are captured by the derivative expansion \eqref{eftofi}. If $c_s\sim \mathcal{O}(1)$, Eq.~\eqref{derexp} ensures  that all the operators $\propto(\bfnabla^{n}\zeta)^2$  with $n\geq 2$ provide subleading corrections to the linearized scalar dynamics. However, it may happen that the system dynamically evolves into a phase in which $c_s\ll 1$. In this case, the  operator  $ (\bfnabla^2\zeta)^2$ may become the leading one, changing  the dispersion relation for $\zeta$.
If this happens before the dynamics becomes strongly coupled, then the system effectively enters a  ghost-condensate-like phase \cite{ArkaniHamed:2003uy,ArkaniHamed:2003uz}. In the following, we will discuss precisely this situation and analyze to what extent this can be used to enhance the power spectrum of $\zeta$ within the EFT \eqref{eftofi}.

\subsection{Linearized mode function equations} \label{basiceqs}

It is convenient to rewrite the linearized equation of motion for $\zeta$ in terms of the number of e-folds $N$, which are related to the cosmological time $t$ and the conformal time $\tau$ through 
\begin{equation}
\D N = H \D t = a H \D \tau \, .
\end{equation}
Defining
\begin{equation}
v\equiv z \zeta \, ,
\qquad
z^2 \equiv \frac{2a^2}{H^2} \left(2 M_2^4 - \MP^2\dot H \right) = 2\MP^2a^2\varepsilon (1+\alpha_1)  \, ,
\label{defzsquared}
\end{equation}
where 
\begin{equation} 
\label{alpha1}
\alpha_1\equiv -\frac{2M_2^4}{\MP^2\dot H}\,,
\end{equation} 
mirroring the notation of \cite{Pirtskhalava:2015zwa}, the quadratic action for $\zeta$ \eqref{ghostL2} can  be rewritten as
\begin{equation}
S_\zeta^{(2)}  = \frac{1}{2 }\int\D\tau\D^3\bfx \left[
 v'^2
- c_s^2 (\bfnabla v)^2
- \alpha \frac{(\bfnabla^2v)^2}{H^2a^2}
+ \frac{z''}{z}v^2
\right] \, ,
\label{ghostL2zeta}
\end{equation}
where the prime ${\,}'$ denotes derivatives with respect to the conformal time $\tau$.
Then, one finds the following Mukhanov--Sasaki type of equation for the mode function $v_k$ in momentum space,
\begin{equation}
v_k'' + \left( c_s^2 k^2 + \alpha  k^4 \tau^2  -\frac{z''}{z} \right) v_k =0 \, .
\label{app:MSeq2}
\end{equation}
Throughout the paper, we will use the following definition for the first two slow-roll parameters:
\begin{equation}
\varepsilon \equiv \frac{\D \log H^{-1}}{\D N} \, ,
\qquad
\eta \equiv \frac{\D \log \varepsilon}{\D N} \, .
\end{equation}
In addition, we define
\begin{equation} \label{sgamma}
s \equiv \frac{\D \log c_s}{\D N} \, ,
\qquad
\gamma_1 \equiv \frac{\D \log (1+\alpha_1)}{\D N} \, .
\end{equation}
We will say that {\it generalized slow-roll}  is satisfied provided that $\varepsilon$, $\eta$, higher order slow-roll parameters and $s$, $\gamma_1$ are $\ll 1$.
With this notation,
\begin{equation}
\frac{z''}{z}	= a^2H^2\left[  \left( 1 +\frac{1}{2}(\eta + \gamma_1 )\right)\left( 2 - \varepsilon +\frac{1}{2}(\eta + \gamma_1 )\right) + \frac{1}{2}\frac{\D (\eta + \gamma_1 )}{\D N} \right] \, , 
\label{zppoz}
\end{equation}
and the equation \eq{app:MSeq2} takes  the form 
\begin{equation}
\frac{\D^2 v_k}{\D N^2 } + (1-\varepsilon) \frac{\D v_k}{\D N } 
+ \left[\left(c_s^2+\frac{\alpha  k^2}{a^2H^2}\right)  k^2-\frac{z''}{z}
\right]\frac{v_k }{a^2 H^2 } =0 \, ,
 \label{MS1}
\end{equation} 
or, equivalently, in terms of $ \zeta_k = v_k/z$,
\begin{equation}
\frac{\D^2 \zeta_k}{\D N^2 } + \left( 3-\varepsilon +\eta + \gamma_1 \right) \frac{\D \zeta_k}{\D N } 
+ \left( \frac{c_s^2 k^2}{a^2 H^2 } +  \frac{\alpha  k^4}{a^4H^4} 
\right) \zeta_k =0 \, .
 \label{MSzeta}
\end{equation}
Setting in \eq{cs} $\bar M_2=0$ (which means $\alpha=0$) and  $\hat M_1=0$ (which implies $1/c_s^2 =1+\alpha_1$, and $\gamma_1=-2s$) we recover the analogous equation studied in \cite{Ballesteros:2018wlw} (setting there $\mu=0$). Later on, we will discuss particular solutions to \eq{MSzeta} and compute the corresponding power spectra, which we will   compare to the slow-roll one. Neglecting for the time being
the  $k^4$-term in the dispersion relation ($\alpha =0$), at  leading order in generalized slow-roll, the power spectrum for $\zeta$ is
\begin{equation}
 \Delta_\zeta^2 \equiv \frac{k^3}{2\pi^2} P_\zeta  = \frac{\left(H/\MP\right)^2}{8\pi^2\,  (1+\alpha_1)\, \varepsilon\, c_s^3}
 \,,
\label{powerspectrumSR}
\end{equation}
which in the aforementioned limit ($\hat M_1=0$) reduces to the usual expression for $c_s\neq1$: 
\begin{align}
\Delta_\zeta^2=\frac{(H/\MP)^2}{8\pi^2 \varepsilon  c_s}\,.
\end{align}

\subsection{Strong coupling}
\label{subsec:strongcoupling}

As in any EFT, a fairly reliable  way  to determine the regime of validity of \eqref{eftofi} is to estimate the energy scale $\Lambda_\star$ at which the  theory becomes strongly coupled or perturbative unitarity breaks down. To this end, one would need to include in \eqref{eftofi} all the  interactions that contribute at the same order in derivatives, e.g.\  the cubic operators  $(\delta g^{00})^3$ and $(\delta g^{00})^2\delta K$ (which  we did not write explicitly in \eqref{eftofi}),  and determine for instance at which energy scale loop corrections become as large as the tree-level diagrams.
As an example, we consider below the operator $\hat{M}_1^3H^{-3}(\bfnabla\zeta)^2\bfnabla^2\zeta$, which results from  $\hat{M}_1^3\delta g^{00}\delta K$. This will be particularly relevant later on, when we consider the case of small $c_s^2$. Indeed, when $c_s\neq 1$, different operators provide in general  different estimates of the strong coupling scale.
In particular, in the limit $c_s\ll1$, the operator $(\bfnabla\zeta)^2\bfnabla^2\zeta$   is the first one to become strongly coupled \cite{Cheung:2007st,Senatore:2009gt,Pirtskhalava:2015zwa}.

Let us start considering the limiting case $\alpha \ll1$, with $c_s^2\lesssim 1$.
If the EFT coefficients are constant in time, or at least do not change too fast---more on this point later---, following e.g.~\cite{Pirtskhalava:2015zwa}, one finds
\begin{equation}
\Lambda_\star^6 \sim (4\pi)^2\MP^2\vert \dot H\vert H^2 \alpha_2^{-2} (1+\alpha_1)^3 c_s^{11} \sim (4\pi)^2 \alpha_2^{-2}(1+\alpha_1)^{3} c_s^{11} \Lambda_3^6   \, ,
\label{scscaleWBG}
\end{equation}
where we have used the expression \eq{Lambda3} for $\Lambda_3$ and we have defined
\begin{equation}
\alpha_2 \equiv -\frac{\hat{M}_1^3H}{\MP^2\dot H} \, .
\end{equation}
It is sometimes convenient to rewrite \eqref{scscaleWBG} in terms of the slow-roll power spectrum \eqref{powerspectrumSR}, 
\begin{equation}
\Lambda_\star^6 \sim  \frac{2H^6 (1+\alpha_1)^2 c_s^{8}}{\alpha_2^2\Delta_\zeta^2}\,.
\end{equation}
Requiring that $\Lambda_\star\gtrsim H$ amounts to the condition
\begin{equation}
(1+\alpha_1)^2 c_s^{8} \gtrsim \alpha_2^2 \Delta_\zeta^2 \, .
\label{condpbh}
\end{equation}
Similar conditions can be obtained from  other interactions in  the EFT \eqref{eftofi}.
Let us assume that initially $\alpha_1, \, \alpha_2$ and  $c_s^2$ are all $\mathcal{O}(1)$.
If  $(1+\alpha_1)$ or $c_s^2$ then  decrease in time, the left-hand side of \eqref{condpbh} becomes smaller, while the power spectrum on the right-hand side increases according to \eqref{powerspectrumSR}. The system approaches therefore strong coupling, resulting eventually in a breaking of the effective expansion once \eqref{condpbh} ceases to hold. When  $c_s^2$ goes to zero, the breakdown of the effective theory can be avoided if, before strong coupling is reached, the operator $\alpha(\bfnabla^2\zeta)^2$ becomes dominant over  $c_s^2(\bfnabla\zeta)^2$ resulting in a change of the dispersion relation \cite{Baumann:2011su}. In this case, the system effectively enters a ghost-condensate phase \cite{ArkaniHamed:2003uy,ArkaniHamed:2003uz}, which provides a weakly-coupled UV completion \cite{Baumann:2011su}. In this new phase, the strong coupling scale, determined by the operator $(\bfnabla\zeta)^2\bfnabla^2\zeta$, is precisely\footnote{Had we chosen the operator $\dot \zeta(\bfnabla\zeta)^2$  instead of $(\bfnabla\zeta)^2\bfnabla^2\zeta$ to  determine the strong coupling scale, we  would have found $\Lambda_\star\sim \Lambda_3/\alpha^2 \gg\Lambda_3$ (assuming again $\alpha_1,\alpha_2\sim\mathcal{O}(1)$) instead of \eqref{scs-ghost}, in agreement with Eq.~(3.25) of \cite{Baumann:2011su}.}
\begin{equation}
\Lambda_\star \sim \Lambda_3 \, ,
\label{scs-ghost}
\end{equation}
where we used \eqref{M2Mb2h} and \eqref{alphaL3}, dropped numerical factors and assumed $\alpha_1,\alpha_2\sim\mathcal{O}(1)$ for simplicity. 

To derive \eqref{scscaleWBG} and \eqref{scs-ghost} we have assumed that all the EFT couplings are (approximately) constant. In general, this is allowed if the typical energy scale associated with the time dependence in the effective couplings, {$\Gamma$}, is {smaller than $H$}. Indeed, the rate of change of some effective coupling $f$  over a Hubble time can be estimated as $H^{-1} \D f/ \D t \sim (\Gamma/H)f$, which will be much smaller than $f$ itself---i.e.\ the variation is slow---provided that $\Gamma \ll H$.
However,   in the following we will be interested in considering situations where   some of the parameters, in particular   $\alpha_1$ and $c_s^2$,  change over time scales {$\Gamma^{-1}$} that are smaller than $H^{-1}$.  Can this affect  the estimate  of  the range of validity of the EFT,  invalidating in particular  \eqref{scscaleWBG} and \eqref{scs-ghost}? 
In principle, these will remain good estimates of the strong coupling scale as long as the typical time scale associated with the time evolution of these coefficients, {$\Gamma^{-1}$,} is larger than $\Lambda_\star^{-1}$.
In the presence of such a scale separation, it is acceptable to estimate the strong coupling scale of the theory---which effectively corresponds to probing the short-distance dynamics of the perturbations in the system---as if the effective coefficients were constant in time. 
In terms of the number of e-folds, requiring that the time scale of a certain `feature' in the effective coefficients or the power spectrum is much larger than $\Lambda_\star^{-1}$ {(which can be thought of as the `time-resolution' of the EFT)} amounts to the condition $\Delta N_{\rm feature} \gg H/ \Lambda_\star$, where we assumed $H\simeq \text{constant}$.
Under this assumption, one can  rely on \eqref{scscaleWBG} and \eqref{scs-ghost} to estimate the regime of validity of the effective theory. The presence of the feature might affect the value of $\Lambda_\star$ by at most an $\mathcal{O}(1)$ correction, but it will not change its order of magnitude.

\section{Enhanced power spectrum from negative friction}
\label{enhanced}

In this section we discuss {one of the possible  mechanisms} by which a large power spectrum can be obtained in the EFT \eqref{eftofi}. This consists in the excitation of a growing mode when it is  outside the horizon because the Hubble friction in the linearized equation for $\zeta$ becomes negative. This mechanism has been studied in the literature when the change of sign in the friction comes from either $\eta$ {(see e.g.\ \cite{Ballesteros:2020qam} and references therein)} or the time variation of the speed of sound {(see \cite{Ballesteros:2018wlw,Ozsoy:2018flq})}. Here we will consider a variation of the EFT coefficient $\gamma_1$---see Eq.\ \eq{sgamma}---which encompasses also the latter of those.


Let us first consider Eq.\ \eq{MSzeta} in the small-$k$ limit. One of the two possible solutions of this equation is simply $\dot\zeta=0$, which (using the appropriate boundary conditions) gives the expression \eq{powerspectrumSR} for the power spectrum if the variation of the EFT coefficients is slow. The other solution satisfies
\begin{equation}
\frac{\D\zeta}{\D N} \propto  \text{exp} \left[-\int (3-\varepsilon + \eta +\gamma_1) \, \D N \right] \, .
\label{dgmode}
\end{equation}
If the sign of the `friction coefficient' $\xi$, defined by
\begin{equation}  \label{friction}
\xi\equiv 3-\varepsilon + \eta +\gamma_1\,,
\end{equation}
becomes negative during inflation, $\zeta$ is not conserved on superhorizon scales and its spectrum may grow significantly above the slow-roll solution \eq{powerspectrumSR}. This friction enhacement of the spectrum can arise from a fast change in $\varepsilon$ which temporarily triggers $\eta < -3$ (see, e.g., \cite{Ballesteros:2017fsr,Ballesteros:2020qam,Ballesteros:2020sre}). We are instead interested in the possibility of $\gamma_1 < -3$, with small and approximately constant $\varepsilon$ and $|\eta|$. From Eq.~\eq{sgamma}  we see that $\gamma_1 <- 3$ can occur if $(1+\alpha_1)$ diminishes  fast enough. In particular, an $\mathcal{O}(1)$ reduction of $\alpha_1$ over one e-fold is necessary for this friction enhancement to take place.  Assuming a sudden change of $\gamma_1$ to a value $\gamma_1<-3$ that remains constant over a time interval $\Delta N$ to decrease rapidly afterwards, the maximum enhancement can be estimated as 
\begin{align}
\Delta_\zeta^2 = \Delta_\zeta^2 (k \rightarrow 0)\,  \e^{-2(3+\gamma_1)\Delta N}\,, 
\label{expenhgamm1}
\end{align}
where  $\Delta_\zeta^2 (k \rightarrow 0)$ is given by \eq{powerspectrumSR}. We emphasize that this effect is different from the enhancement that can happen from a slow decrease of $(1+\alpha_1)$, which is already explicit in \eq{powerspectrumSR}. 

The exponential enhancement \eqref{expenhgamm1}  is in principle possible, but we should check that it does not violate perturbative unitarity in the EFT, nor any of the assumptions in Section~\ref{subsec:strongcoupling}.
In particular, we will require that \eqref{scscaleWBG} is valid and that the inequality \eqref{condpbh} is satisfied. For simplicity, let us start assuming $c_s$ constant---we shall discuss later cases where $c_s$ evolves in time.  Plugging \eqref{powerspectrumSR} into  \eqref{condpbh} and assuming $\alpha_1, \, \alpha_2\sim \mathcal{O}(1)$ initially, one infers that  $(1+\alpha_1)$ cannot get smaller than roughly $\sim 10^{-3}$.
In other words, defining $1+\alpha_1\equiv 10^{-y}$, weak coupling requires roughly $\Delta y \lesssim 3$, if $y$ is initially close to zero.
On the other hand, in order for the friction coefficient \eqref{friction} to become negative, one needs $\Delta y \gtrsim 3 \Delta N /\log 10$. Combined with $\Delta y \lesssim 3$, this implies  $\Delta N \lesssim2$. However, from the considerations in Section~\ref{subsec:strongcoupling}, we want $\Delta N \gtrsim \mathcal{O}(1)$ to trust \eqref{scscaleWBG} in the first place. Therefore, one concludes that, even if it is certainly possible to obtain a significant enhancement of the power spectrum by means of a large and negative $\xi$, this may require some additional assumptions about the UV physics. Otherwise,  based on   the results of Section~\ref{subsec:strongcoupling}, the enhancement is bounded by the combined  requirements that $(1+\alpha_1)$ becomes not too small and its variation is not too fast.

\begin{figure}
  \includegraphics[scale=0.55,trim=2.1cm 13cm 1cm 4cm,clip]{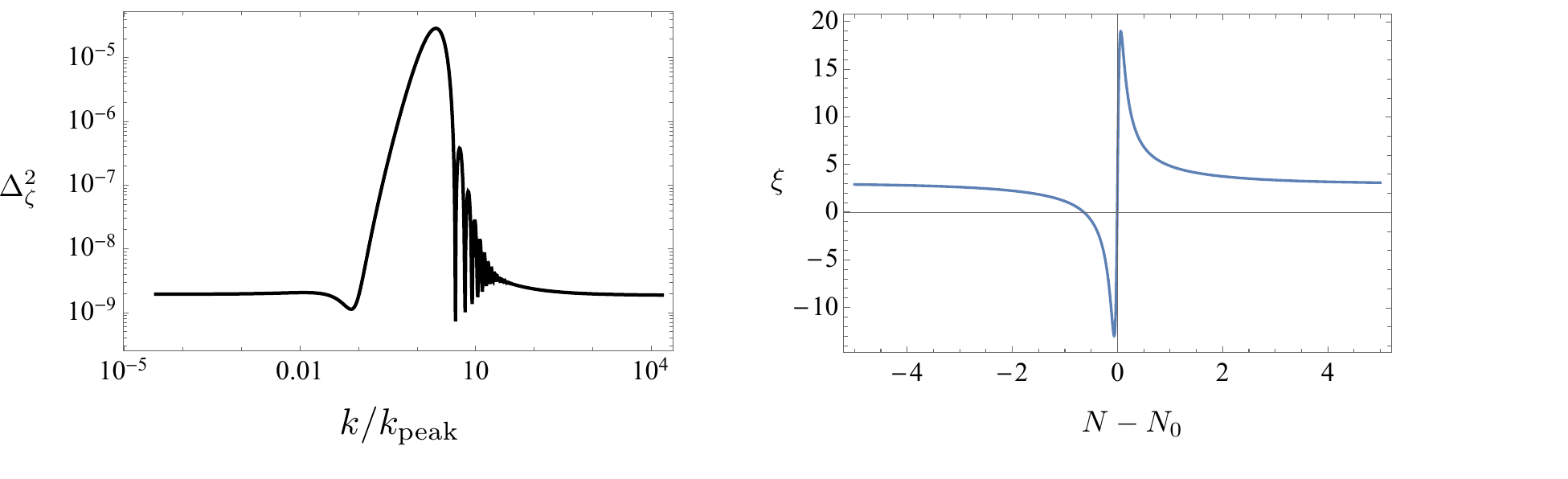}
\caption{We show the numerical power spectrum as a function of the momentum $k$,  given the following time dependence in the parameter $\alpha_1$: $1+\alpha_1= (1-\alpha_f)\tanh(w(N-N_0))^2+\alpha_f$, where  $\alpha_f=10^{-3}$ and $w=1/3$. We have assumed  $c_s^2=1$ and  $\varepsilon=10^{-3}$. $N_0$ denotes the  (arbitrary) e-fold time around which the transition happens.}
\label{Fig:a1}
\end{figure}

We consider an explicit example in  Figure~\ref{Fig:a1}, where we show the power spectrum
for a particular choice of the time evolution of the coefficient $\alpha_1$. In the figure,  $(1+\alpha_1)$ decreases from an $\mathcal{O}(1)$ number by roughly $3$ orders of magnitude, in a time lapse of a couple of e-folds, before increasing again back to its original value.\footnote{In this example $c_s$ is constant for simplicity. This is possible in spite of the change in  $M_2$ (required to have a variation of $\alpha_1$), thanks to the freedom in the choice of $\hat M_1^3$, see Eqs.\ \eq{cs}.} As a result, this translates into an enhancement of the power spectrum of about four orders of magnitude, 
with respect to the CMB value, which has been set at $k\ll k_\text{peak}$ to be $\sim 2\times10^{-9}$.

It is instructive to compare the previous case to the one  in which $\hat M_1=0$. 
As we mentioned at the end of Section \ref{basiceqs}, if $\hat M_1=0$ then $1/c_s^2 = 1+\alpha_1$ and $\gamma_1 = -2 (\D \log c_s/\D N)=-2s$. In this case, assuming a slow variation of the EFT coefficients, $\Delta_\zeta^2\propto 1/c_s$ and a reduction of $c_s$ enhances the slow-roll spectrum. However,  $\gamma_1>0$ when $\D c_s/\D N<0$ (assuming $c_s>0$) and such a variation cannot enhance the spectrum by means of the friction effect discussed above, as it only makes the non-constant mode of $\zeta$ fall even more rapidly as $N$ grows. The enhancement of the power spectrum from a change in $c_s^2$ was discussed in \cite{Ballesteros:2018wlw} in analogous terms. There, it was pointed out that a small sound speed, leading to a large $\Delta_\zeta^2\propto 1/c_s$ (as required for abundant PBH formation) can lead to strong coupling and the loss of validity of the effective theory.
In that case, the strong coupling condition \eqref{condpbh} reduces indeed to $c_s^4 \gtrsim \Delta_\zeta^2$ \cite{Creminelli:2006xe}, which provides a lower bound on $c_s$. In the following, we  we will explore further  the possibility of  enhancing the power spectrum through a change in  the sound speed, as well as possible ways to avoid strong coupling issues.

\section{Enhanced power spectrum from a ghost-inflation phase}
\label{sec:srtogis}

In Ref.\ \cite{Ballesteros:2018wlw} it was pointed out  in the context of PBH formation that the strong coupling problem that arises when one tries to enhance the power spectrum to large values by sending  $c_s\rightarrow 0$ may be addressed by considering a modified dispersion relation in this limit, and specifically the case of ghost inflation \cite{ArkaniHamed:2003uz}.\footnote{Another possibility consists in invoking the appearance of new degrees of freedom that provide a UV completion to the effective theory, see Appendix~\ref{app:2-fields}.} This is because a change in the dispersion relation for the low-energy degrees of freedom can raise the cutoff of the theory, extending the regime of validity of the effective description. In the following, we discuss  in detail  this possibility. In particular, we  consider the case in which the system transits from a slow-roll regime into a  ghost-condensate-like phase.  This is possible provided that $\hat M_1^3$ is appropriately chosen. 
The robustness of the transition is ensured by the assumed WBG  symmetry \cite{Pirtskhalava:2015nla,Pirtskhalava:2015zwa}.
{
As we will see, this type of transition, which modifies the dispersion relation for the perturbations, allows to simultaneously $i)$ enhance the power spectrum on shorter scales and $ii)$ keep the theory for the perturbations weakly coupled during the evolution.
}

The initial slow-roll evolution is characterized by the usual dynamics (with $c_s\sim \mathcal{O}(1)$ initially),
\begin{equation}
v_k'' + \left( c_s^2 k^2   -\frac{2}{\tau^2} \right) v_k =0 \, ,
\label{eqsrst}
\end{equation}
which admits the standard solution
\begin{equation}
v_k^{\text{SR}}(\tau) = \frac{\e^{-ic_sk\tau}}{\sqrt{2c_sk}} \left( 1- \frac{i}{c_sk\tau} \right) \, ,
\label{m2}
\end{equation}
where the Wronskian condition $v_kv_k'^*-v_k'v_k^*=i$ has been imposed at $\tau\rightarrow-\infty$.
As $c_s$ decreases in time,  the slow-roll power spectrum \eqref{powerspectrumSR}, obtained from the solution \eqref{m2}, increases, but at the same time  the system approaches strong coupling. However, before it becomes strongly coupled,  we   assume that the dynamics of the perturbations gets dynamically modified, in particular that  the term proportional $\alpha k^4$ becomes dominant over $c_s^2 k^2$ in  Eq.\ \eqref{app:MSeq2}:
\begin{equation}
v_k'' + \left( \alpha  k^4 \tau^2  -\frac{2}{\tau^2} \right) v_k =0 \, .
\label{ghostphasemodeeq}
\end{equation}
Given a certain mode with momentum $k$, the crossover between the two regimes occurs at the conformal time $\tau_*$, defined by  $c_{s}(\tau_*) \sim \sqrt{\alpha(\tau_*)} (- k \tau_*) $. 
{Modes with very small $k$ that exit the horizon very soon will not be affected by the change in the dynamics, and will  have the power spectrum given in \eqref{powerspectrumSR}.
Let us focus instead on modes with larger $k$, that  were still sub-horizon during the transition into the ghost-inflation like phase. In order to estimate the power spectrum for these modes, we  can proceed as follows. 
First, we solve the mode function equation \eqref{ghostphasemodeeq}. The most general solution can be written as  \cite{ArkaniHamed:2003uy,ArkaniHamed:2003uz}
}
\begin{equation}
v_k^{\text{GI}}(\tau) = A_k \sqrt{-\tau} H^{(1)}_{3/4}\left(\frac{1}{2}\sqrt{\alpha} \, k^2 \tau^2 \right) 
+ B_k \sqrt{-\tau} H^{(2)}_{3/4}\left(\frac{1}{2}\sqrt{\alpha} \, k^2 \tau^2 \right)  \, .
\label{solNR}
\end{equation}
Next, we  match \eqref{solNR} to \eqref{m2} by requiring continuity across the transition. This will unambiguously fix  the coefficients $A_k$ and $B_k$ in \eqref{solNR}.
Since the modes  which we want to estimate the power spectrum for after the transition were well within the horizon in the  slow-roll phase, we can obtain a reasonable approximation to the solution in the ghost-condensate phase by matching \eqref{m2} and \eqref{solNR} at large negative conformal times, $\tau=-\infty$.\footnote{{Clearly the correct procedure would require matching $\zeta $ and $\zeta'$ at the crossing point---see, e.g.\ Figure~\ref{fig:SRtoGI} below. However, doing this does not  significantly affect the order of magnitude of the enhancement in \eqref{srovergipw}.}} Using the following formulae for the Hankel functions, 
\begin{equation}
H_p^{(1)}(z\rightarrow+\infty) \approx  \sqrt{\frac{2}{\pi z}} \e^{i\left( z- p\frac{\pi}{2} - \frac{\pi}{4} \right)} \, ,
\qquad
H_p^{(2)}(z\rightarrow+\infty) \approx  \sqrt{\frac{2}{\pi z}} \e^{-i\left( z- p\frac{\pi}{2} - \frac{\pi}{4} \right)} \, ,
\label{idhankel}
\end{equation}
and keeping only the solution proportional to $A_k$ in \eqref{solNR}, which has the correct phase at short distances, we can approximate
\begin{equation}
v_k^{\text{GI}}(\tau\rightarrow -\infty ) \approx A_k \, \e^{-i\frac{5\pi}{8}}  \frac{2}{\alpha^{1/4} k \sqrt{-\tau} \sqrt{\pi }}  \e^{ \frac{i}{2}\sqrt{\alpha} \, k^2 \tau^2} \, .
\label{solNR2}
\end{equation}
By definition, at the transition one has $c_s(\tau_*)\sim \sqrt{\alpha_*}  (-k\tau_*)$. Then, 
\begin{equation}
v_k^{\text{GI}}(\tau\approx\tau_*) \approx  \frac{2\sqrt{2} \, A_k\, \e^{-i\frac{5\pi}{8}+ \frac{i}{2}c_s(\tau_*)k \tau_*} }{\sqrt{\pi }} \frac{\e^{- ic_s(\tau_*)k \tau_*} }{ \sqrt{2kc_s(\tau_*)} }  \, ,
\label{e1}
\end{equation}
and the matching with \eqref{m2}  yields
\begin{equation}
A_k = \frac{\sqrt{\pi }}{2\sqrt{2}  }\e^{i\frac{5\pi}{8}- \frac{i}{2}c_{s*}k \tau_*} \, .
\label{Akmatch}
\end{equation}
Expanding at $\tau\rightarrow0$, one finds the following estimate for the power spectrum after the transition,\footnote{We stress that   \eqref{p1} holds only in the limit of very large $\vert \tau_*\vert$. Keeping $\tau_*$ finite   introduces corrections that depend on $c_s$ and $\alpha $ computed at $\tau_*$. A more precise realization of the transition is discussed later, with the resulting power spectrum shown in Figure~\ref{fig:SRtoGI}.  }
\begin{equation}
\Delta_{\zeta,\text{GI}}^2  
 \approx \frac{ H^4 }{2 \pi \, \Gamma(\frac{1}{4})^2 \, \alpha^{3/4} \left(2 M_2^4 - \MP^2\dot H \right) }= 
\frac{ \left(H/\MP\right)^2 }{2 \pi \, \Gamma(\frac{1}{4})^2 \,  \alpha^{3/4} \left(1+\alpha_1 \right)\varepsilon	} 
  \, ,
 \label{p1}
\end{equation}
to be compared with the slow-roll power spectrum \eqref{powerspectrumSR}, which gives
\begin{equation}
\Delta_{\zeta,\text{GI}}^2  \simeq \frac{4\pi}{ \Gamma(\frac{1}{4})^2 }\,\frac{c_s^3}{\alpha^{3/4}}\, \Delta_{\zeta}^2(k\rightarrow 0)\,,
\label{srovergipw}
\end{equation}
where $c_s$ is the  sound speed in the slow-roll phase before the sound speed starts changing in time. 
If we take $c_s\sim \mathcal{O}(1)$ before the transition, then it follows from \eqref{srovergipw} that the final power spectrum is enhanced by a factor of $\alpha^{-3/4}$ (recall that $\alpha\ll1$) for modes that exit the horizon after $\tau_*$, with respect to modes that become super-horizon during the slow-roll phase.

In order to understand the origin of the enhancement, we shall reason as follows. It is convenient to focus on the two-point functions resulting from the solutions \eqref{m2} and \eqref{solNR} as functions of $k\tau$. In particular, let us assume for simplicity that $c_s$ is the only time-dependent quantity, with $c_s=1$ initially, while all the other parameters are constant, and let us define the functions $\Delta_{1}(z)\equiv c_s^{-1}(z^2+c_s^{-2})$ and $\Delta_{2}(z)\equiv 2 \vert A_k\vert^2 (-z)^3\vert H^{(1)}_{3/4}\left(\frac{1}{2}\sqrt{\alpha} \, z^2 \right) \vert^2$, where $z\equiv k \tau\leq0$. $\Delta_{1}$ is related to the power spectrum \eqref{powerspectrumSR} in the slow-roll phase via $\Delta_{\zeta,\text{SR}}^2= \frac{\left(H/\MP\right)^2}{8\pi^2\,  (1+\alpha_1)\, \varepsilon} \Delta_{1}(z\rightarrow0)$, while $\Delta_{2}$ gives, in the limit $z\rightarrow0$, the power spectrum \eqref{p1} in the ghost-condensate phase after multiplying by the same overall constant and using \eqref{Akmatch}.
As already mentioned above, modes with momentum $k$ that exit the horizon well before the sound speed starts decreasing   will not be affected by the change in the dynamics and will have a  power spectrum  determined by $\Delta_{1}(z\rightarrow0)$ with $c_s=1$.
Instead, the power spectrum of modes that  become super-horizon while $c_s$ is decreasing (but before their dispersion relation switches from linear to quadratic) is still determined by $\Delta_{1}$ but with a smaller value for $c_s$.
In particular, the  enhancement of these modes with respect to the previous ones scales (in the slow-roll approximation) as $c_s^{-3}$ at horizon crossing, as dictated by  Eq.~\eqref{powerspectrumSR}. 
 Finally, let us consider modes with values of $k$ that exit the horizon in the ghost-condensate phase. These modes are still sub-horizon around $\tau=\tau_*$. Before $\tau_*$,  their amplitude follows (up to an overall constant factor) the curve $\Delta_1$ as a function of $k\tau$, while after $\tau_*$ they will evolve according to $\Delta_2$. However,
since $\Delta_1$ and $\Delta_2$ have different shapes in $z$, and, in particular, since $\Delta_2$ decreases more slowly  than $\Delta_1$, the  power spectrum  is larger than what it would be without the change in the mode dynamics. The enhancement   scales precisely as  $\alpha^{-3/4}$. 
We emphasize that this precise scaling $\alpha^{-3/4}$ holds strictly speaking in the limit of very large $\vert \tau_*\vert$. Considering a shorter duration for the ghost-inflation-like phase will in general translate into a smaller enhancement. 
In addition, the actual enhancement in the power spectrum will also depend  on the details of the transition and on the presence of oscillations,  which may result in a larger peak.
We show an explicit example in Figure~\ref{fig:SRtoGI}, where we plot $\Delta_\zeta^2(k)$.
In the left panel, we have assumed that the system is initially  in a slow-roll phase with $c_s^2=1$ and $\alpha$ negligible. At some time in the evolution, $c_s^2$ goes to zero and the dynamics becomes thereafter dominated by the $k^4$-term.\footnote{In reality, $c_s$ will never be exactly zero. It is indeed bounded from below by the size of the quantum corrections to the effective couplings in \eqref{cs}.    }
As shown by the  orange dashed line, the slope of the spectrum goes as $\sim k^4$. Note that this is the same slope that was found in \cite{Byrnes:2018txb}, although the growth there is due to a change in the background evolution (in particular, a transition into an ultra-slow-roll phase) instead of a change in the dispersion relation for the perturbations.  
In the right panel, the evolution is the same except that the phase with quadratic dispersion relation has finite duration:  $\sim 6$ e-folds after the first transition, the sound speed becomes $c_s^2=1$ again, and the initial  slow-roll dynamics   is restored.  To make the analysis as simple as possible, we have assumed the transitions between the different phases instantaneous, and we have matched the solutions for  $\zeta $ and $\zeta'$ at the transition points.
\begin{figure}[h!] 
\begin{center} 
\includegraphics[scale=0.22]{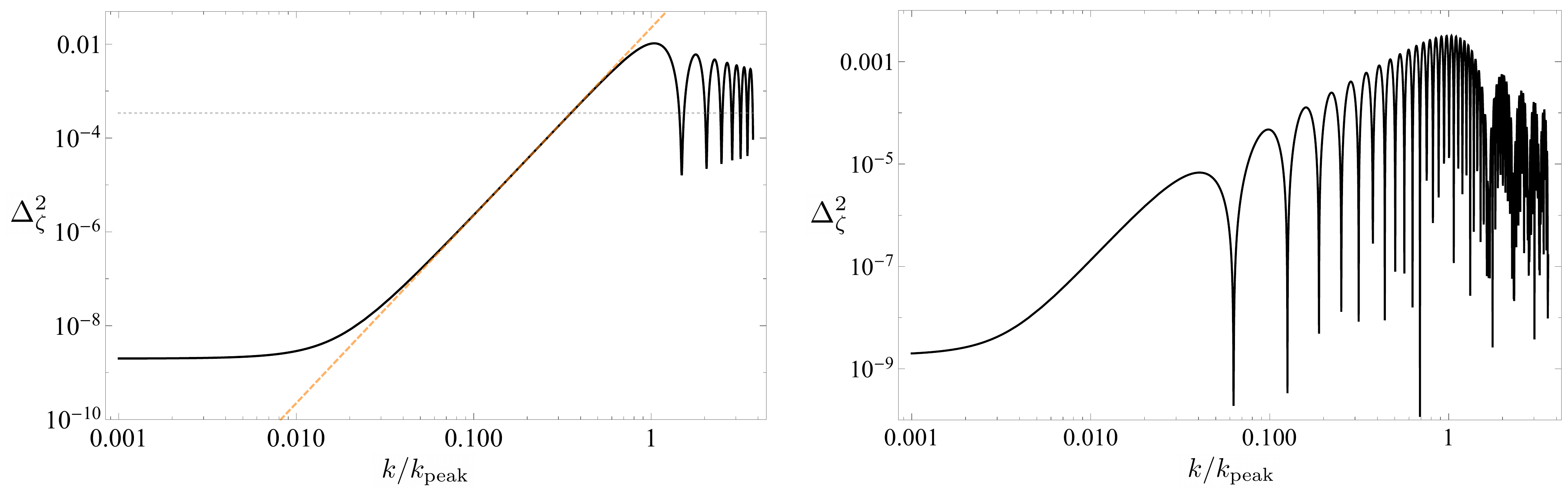}
\caption{  Power spectrum, 
$\Delta_\zeta^2(k)$, normalized to $\Delta_{\zeta}^2(k\rightarrow 0) = 2 \times 10^{-9}$.
\underline{Left panel:} We assume a transition between a slow-roll phase with $c_s=1$ and $\alpha $ negligible, and a phase with quadratic dispersion relation ($c_s=0$ and $\omega = \sqrt{\alpha}\,  k^2\tau^2$, for $\alpha=10^{-7}$). 
The horizontal dashed line is given by Eq.~\eqref{srovergipw}. The orange dashed line has the slope $k^4$.
The figure illustrates that the presence of the oscillations allow to get a peak whose actual amplitude is between $1$ and $2$ orders of magnitude larger than the value estimated with Eq.~\eqref{srovergipw}.
\underline{Right panel:} We assume now  that  the phase with quadratic dispersion relation has a finite duration ($\Delta N\simeq 6$). After that, the initial dynamics with $c_s^2=1$ is restored. 
In the intermediate phase,   $\alpha=10^{-10}$ in this case.    }
\label{fig:SRtoGI}
\end{center} 
\end{figure}

Looking at Eq.~\eqref{srovergipw}, one might  be tempted to conclude that, by suitably choosing $\alpha$ to be sufficiently small (which, using the scaling \eqref{alphaL3}, would correspond to raising the scale $\Lambda_3$)
it is possible to obtain {an arbitrarily}  large relative enhancement 
of the 
power spectrum with respect to the standard slow-roll one. 
Nevertheless, we 
know that $\alpha$ cannot be arbitrarily small (or, $\Lambda_3$  arbitrarily large---see Appendix~\ref{app:sizeoperators} for further details). Indeed, a nonzero $\alpha$ is     generated anyway  at the quantum level.
Then, the relevant question is whether the possible values of $\alpha$ that are compatible with the size of its quantum corrections allow to obtain the desired enhancement.
Note that, according to \eqref{srovergipw},  a relative increase of roughly $6$ or $7$ orders of magnitude would require $\alpha$ of order $\sim 10^{-9}$. This seems hard to get using the  values in Table~\ref{tab1} in Appendix~\ref{app:sizeoperators} for the effective couplings (even in the kinetically-driven scenario)   and the typical values of $H$ during inflation. However, one might 
get {such small} numbers for $\alpha$ if the parametric separation between $M_2^4$ and $\bar{M}_2^2H^2$ is larger than the one we  considered here (see Eq.~\eqref{M2Mb2h}), which would correspond to further suppressing the relative size of the higher derivative operators with respect to standard $P(\phi,(\partial_\mu \phi)^2)$ operators in the theory. 
However, as mentioned above, oscillatory features such as those displayed in Figure \ref{fig:SRtoGI} may lead to a larger spectrum (possibly by a couple of orders of magnitude) for a given $\alpha$, depending on the specific implementation of the transition into the ghost-condensate-like phase.

To recap,  a phase  with a quadratic dispersion relation {($\omega\propto k^2$)} allowed us to attain two main results at the same time: firstly, it raised the cutoff of the theory, avoiding short modes (with {large} $k$) to become strongly coupled as $c_s\rightarrow0$; secondly, it resulted in an enhanced power spectrum at large $k$, compared to the power spectrum at low $k$, thanks to a different time evolution for the modes in the  slow-roll  and ghost-inflation phases.
The enhancement scales as $\alpha^{-3/4}$, with $\alpha$ bounded from below by the size of the quantum corrections in the theory. In the case of interest, the typical value of $\alpha$ can be read off Eq.~\eqref{alphaL3} and Table~\ref{tab1} in Appendix~\ref{app:sizeoperators}.

\section{Transition into a phase with imaginary sound speed} \label{ima}
\label{sec:imaginaryspeed}

\subsection{Exponential growth from ghost inflation}
\label{sec:gi-to-ics}

A transition into a ghost-inflation-like phase allowed us  to alleviate the strong coupling regime that stems from reducing  the sound speed  of the perturbations and, at the same time, to enhance the power spectrum on short scales. However, since the relative increase is a power law in $\alpha$ (in particular scales as $\alpha^{-3/4}$),  an enhancement of seven orders of magnitude  (as required by the Gaussian approximation to account for all DM with PBHs formed during radiation domination)\footnote{{We reiterate that this approximation is unlikely to be precise given that it may not describe well enough the tails of the probability distribution function of $\zeta$. Still, a large power spectrum of $\zeta$ is certainly needed for a large PBH abundance.}} 
 is only possible at the price of  further increasing  the  parametric separation between $H$ and $\Lambda_3$, which is tantamount to pushing the form of the theory toward a standard $P(\phi,(\partial_\mu \phi)^2)$ theory.
A different, and in principle  more efficient, way to enhance the power spectrum is to excite a growing mode by assuming that, after transiting into the ghost-condensate phase, the quantity  $c_s^2$ in \eqref{cs} keeps decreasing and eventually becomes negative.
This corresponds to an instability for the perturbation $\zeta$, which now grows exponentially. 
Usually, this is  an unwanted behavior for the perturbations, as this sort of instability heralds
a breakdown of the perturbative expansion and  threatens the validity of the background  classical solution. In the following, we will discuss to what extent such  a gradient instability can be used to enhance the power spectrum in a controlled way over a period of time of a few e-folds.
Caveats and possible obstructions are discussed later in this section and in Section~\ref{sec:ng-pt}.

We remain agnostic about the UV mechanism that is responsible for the exponential growth of the perturbations in the effective theory. In particular, we do not rely on any specific multi-field model.\footnote{In \cite{Fumagalli:2019noh,Bjorkmo:2019qno,Ferreira:2020qkf,Garcia-Saenz:2018ifx} (see also \cite{Fumagalli:2020adf,Fumagalli:2020nvq} and Appendix \ref{app:2-fields}) a gradient instability arises as a consequence of a transient tachyonic instability in a two-field model. In these works, a heavy degree of freedom becomes temporarily tachyonic. Once integrated out, the tachyonic phase manifests as a gradient instability in the infrared dynamics. }
In our analysis, the transition from the slow-roll phase into a transient regime of exponential growth is entirely captured by the weakly coupled, single-field effective theory \eqref{eftofi} and it is driven by  higher derivative operators with WBG symmetry  \cite{Pirtskhalava:2015nla,Santoni:2018rrx}. The robustness of the classical trajectory and  of the EFT is guaranteed by the approximate  symmetry, which protects the effective couplings against large quantum corrections.\footnote{{See also \cite{Garcia-Saenz:2018vqf} for a multi-field motivated EFT description of the gradient instability, albeit without 
any discussion about the role of the higher derivative operators (responsible for a modified dispersion relation at large momenta) or any symmetry considerations.}}

For the moment, let us start by focusing on the transition between the ghost-condensate regime and the phase with imaginary sound speed. Later on, we will consider a more complete scenario, in which these are  transient phases in a more general inflationary slow-roll evolution.

In the ghost-condensate phase, the sound speed is effectively negligible and the linearized dynamics for the perturbations is described by \eqref{ghostphasemodeeq}. After this phase, we assume that $c_s^2$ evolves from zero to  negative values.   The dynamics is  now governed by the mode equation 
\begin{equation}
v_k'' + \left(- c_s^2 k^2  +\alpha k^4\tau^2 -\frac{2}{\tau^2} \right) v_k =0 \, , \label{eq:tgi}
\end{equation}
which corresponds to  \eqref{app:MSeq2} at the leading order in the quasi-de Sitter  approximation and where we replaced, for convenience, $c_s\mapsto i c_s$ in such a way to work with positive $c_s^2$. In general, the  equation \eqref{eq:tgi} cannot be solved exactly in the presence of a non-trivial time dependence in $c_s^2$ and $\alpha$. Therefore, we will take them constant and approximate the evolution  with a sharp transition from a ghost-condensate phase (with $c_s=0$ and $\alpha=\text{constant}$) to a gradient-instability phase (with $\alpha=0$ and $\vert c_s\vert=\text{constant}$).  We will thus solve the equation  separately in the two phases, and  match the solutions at the crossing point.
To this end, let us 
define  $\tau_0\equiv -\vert c_s\vert/(k\sqrt{\alpha})$ to be the (conformal) time when the transition takes place.
In the ghost-condensate phase,  $\tau\ll\tau_0$, the $k^4$ term dominates in \eqref{eq:tgi} and the solution is given by 
\begin{equation}
v_k(\tau)=\sqrt{\frac{\pi}{8}}\sqrt{-\tau} \, H^{(1)}_{3/4}\left(\frac{1}{2}\sqrt{\alpha} \,  k^2\tau^2\right) \, ,\label{eq:GIcurvature}
\end{equation}
where the standard boundary condition for ghost inflation has been imposed at $\tau\rightarrow-\infty$ \cite{ArkaniHamed:2003uz}.
After the transition time $\tau_0$, the $k^2$ term dominates in \eqref{eq:tgi} and the most general solution for $v_k$ takes the form
\begin{equation}
v_k(\tau)=\sqrt{-\tau}\left(A_k H_{3/2}^{(1)}(-i c_s k \tau)+B_k H_{3/2}^{(2)}(-i c_s k \tau)\right) \, .
\label{eq:instabGI}
\end{equation}
The coefficients $A_k$ and $B_k$ can be 
obtained by matching to \eqref{eq:GIcurvature}, requiring the continuity of $\zeta_k\equiv v_k/z $ and $\zeta_k'$ across the transition point:
\begin{align}
A_k&= \frac{(-1)^{1/4} \pi  \e^{\frac{c_s^2}{\sqrt{\alpha }}} }{8 c_s \alpha^{1/4}} \left[\left(c_s^2-\sqrt{\alpha}\right)H_{-1/4}^{(1)}\left(\frac{c_s^2}{2 \sqrt{\alpha }}\right)- c_s^2 H_{3/4}^{(1)}\left(\frac{c_s^2}{2 \sqrt{\alpha }}\right)\right]
 \, ,
\\ 
B_k&=  -\frac{(-1)^{1/4} \pi  \e^{-\frac{c_s^2}{\sqrt{\alpha }}} }{8 c_s \alpha^{1/4} }
\left[\left(c_s^2+\sqrt{\alpha}\right) H_{-1/4}^{(1)}\left(\frac{c_s^2}{2 \sqrt{\alpha }}\right)+c_s^2 H_{3/4}^{(1)}\left(\frac{c_s^2}{2 \sqrt{\alpha }}\right)\right]
 \, ,
\end{align}
where we  used $\tau_0\equiv -\vert c_s\vert/(k\sqrt{\alpha})$.
These expressions can be further simplified  using $\sqrt{\alpha}/\vert c_s\vert^2 \ll1$. 
Expanding the Hankel functions as in \eqref{idhankel} for $c_s^2/(2\sqrt{\alpha})\gg1$, we find
\begin{align}
\label{A}
A_k& \approx   \sqrt{\frac{\pi }{8}} (-1)^{3/8} \e^{\frac{ c_s^2}{\sqrt{\alpha }}+ \frac{i  c_s^2}{2\sqrt{\alpha }} }
 \, ,
\\ 
\label{B}
B_k& \approx  i  \sqrt{\frac{\pi }{8}} (-1)^{3/8} \e^{-\frac{ c_s^2}{\sqrt{\alpha }}+ \frac{i  c_s^2}{2\sqrt{\alpha }} }
 \, .
\end{align}
Plugging these expressions back into \eqref{eq:instabGI}, the 
power spectrum for $\zeta$ is
\begin{equation}
\Delta_\zeta^2 \equiv \frac{k^3}{2\pi^2} \vert \zeta_k\vert^2  \approx \frac{H^2 \e^{\frac{2c_s^2}{\sqrt{\alpha}}}}{16\pi^2 \MP^2 \varepsilon (1+\alpha_1) c_s^3}
 \, .
\label{psenh1}
\end{equation}
where we used \eqref{defzsquared} and neglected subleading terms in $\sqrt{\alpha}/c_s^2$.
{Eq.~\eqref{psenh1} can also be written as}
\begin{equation}
\Delta_\zeta^2 
  \approx     \frac{\pi \alpha^{3/4}  \e^{\frac{2c_s^2}{\sqrt{\alpha}}}  }{4 c_s^3 \Gamma\left(\frac{3}{4} \right)^2}
  \Delta_{\zeta}^2 (k\rightarrow0)
 \, ,
\label{psenh-2}
\end{equation}
where  $ \Delta_{\zeta}^2 (k\rightarrow0)  \simeq 2 \times 10^{-9}$ {(normalized to the CMB measurements)} corresponds to the power spectrum of the modes that  exited the horizon before the instability phase (i.e.\ during the ghost-inflation phase).
From \eqref{psenh-2}  one can thus read off the enhancement factor $\pi \e^{2x} / [4 x^{3/2}  \Gamma(\frac{3}{4} )^2]$, where we defined $x\equiv c_s^2/\sqrt{\alpha}$.
In principle, it is enough to choose $c_s$ and $\alpha$
in such a way that $x\simeq 10$ to obtain an { enhanced spectrum} as large as $\simeq 10^{-2}$.
Note that the power spectrum \eqref{psenh-2} is analogous to the one that arises from the tilted version of  ghost inflation \cite{Senatore:2004rj}. Yet, there are differences worth remarking.
As opposed to \cite{Senatore:2004rj}, where the growth of the perturbations stems 
 from an increasing $H$ in time, which violates the Null Energy Condition (NEC), in our case the background evolution is not altered  with respect to standard slow-roll inflation, i.e.\ with  $\dot{H}<0$ and $-\dot{H}/H^2\ll1$. Instead, the instability that leads to a growth of the primordial perturbations is triggered
by higher derivative operators in the underlying 
theory with WBG symmetry, resulting in a temporary change of the sign of $c_s^2$ (which we have approximated as an instantaneous transition).

It is worth stressing that \eqref{psenh-2} gives the maximum enhancement of the power spectrum. Recall indeed that we have confined our discussion so far to a single mode $k$, while one should note that  not all the modes experience the same relative enhancement. This will be transparent in the next section (see, in particular, Figure~\ref{fig:TransitionPNP-L}), where we consider an analytic approximation for the transient unstable phase and analyze in more detail the behavior of the different modes during the evolution.

\subsection{Analytic approximations for the  transition into the unstable phase}
\label{app:rsexpenh}

In this section, we consider a more refined version of the dynamics we just discussed above. In particular, we assume that the system evolves from a slow-roll phase into a transient unstable phase dominated by a dynamics for the perturbations with $c_s^2<0$. Then, after some time,  $c_s^2$ turns positive again and the system recovers a slow-roll type of evolution. To make this scenario tractable analytically, we approximate the evolution by three separate phases, connected by instantaneous transitions, where the solutions to the equations of motion for the perturbations are matched. This will allow us to show explicitly the behavior of the different modes. A smoother version of the evolution is discussed in the next section, but all the main  ingredients and qualitative aspects are already captured by the present stepwise approximation.

\begin{figure}[h!] 
\begin{center} 
\includegraphics[scale=0.35]{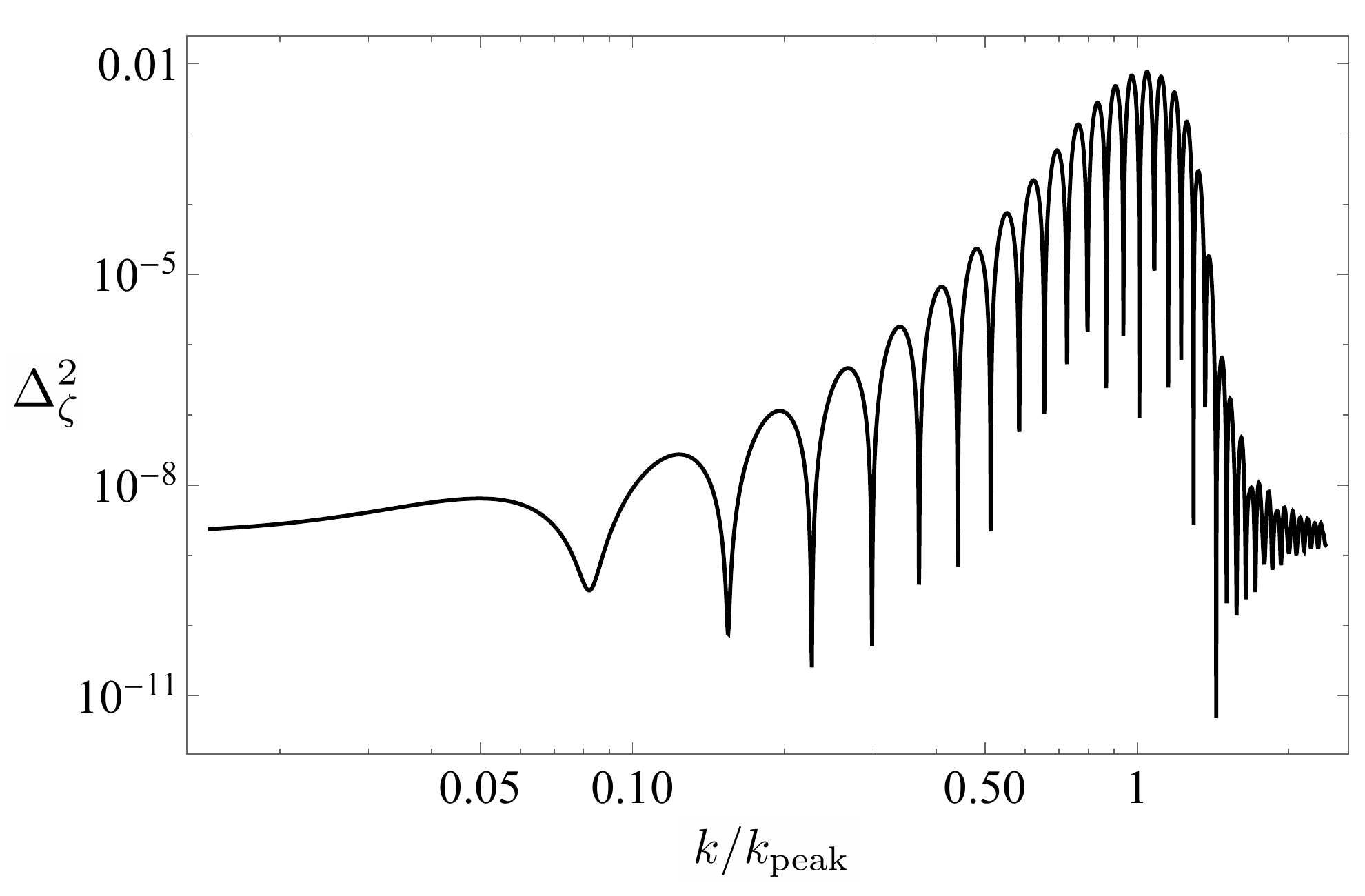}
\caption{
Power spectrum of $\zeta$ as a function of $k$ for  $\bar{c}_s^2=1$,  $\tilde{c}_s^2=0.55$, $\alpha=10^{-4}$, $\tau_t k_\text{peak} = - 57.75$ and $\tau'_t k_\text{peak}  = - 44$ (see Section \ref{app:rsexpenh}). Modes with $k/k_\text{peak}\ll1$ that exit the horizon well before $\tau_t$ are not affected by the gradient instability ($c_s^2<0$). The first modes that start feeling its effect are those with $k/k_\text{peak}\gtrsim0.02$. Modes with larger, intermediate, $k$ are those that are enhanced the most. Those with even larger $k$ ($k/k_\text{peak}>1$) are
instead 
suppressed {(with respect to the peak)}, as the $k^4$-term in \eqref{eq:tgi} starts competing with the $\tilde{c}_s^2k^2$ term, mitigating the effect of the instability. Modes with very large values of $k/k_\text{peak}$, not displayed in the plot, are beyond the effective expansion, as they violate \eqref{derexp}.}
\label{fig:TransitionPNP-L}
\end{center} 
\end{figure}

Let us start assuming that $c_s(\tau)$ is the only time-dependent function determining the dynamics of the perturbations, while  we will keep for simplicity  all the other slow-roll parameters, including $\alpha$ in  \eqref{eq:tgi}, constant.
To make the analysis  as simple as possible, we will assume that $c_s^2$ changes instantaneously from a certain constant value to another  constant value. In particular, we shall define $\tau_t$ and $\tau'_t$ such that $c_s^2(\tau<\tau_t)=\bar{c}_s^2>0$, $c_s^2(\tau_t<\tau<\tau'_t)=-\tilde{c}_s^2<0$ and $c_s^2(\tau>\tau'_t)=\bar{c}_s^2>0$, where $\bar{c}_s^2\simeq1$ and $\tilde{c}_s^2\ll1 $.
In the first phase defined by $\tau<\tau_t$, the scalar perturbations have a standard 
{dispersion relation with 
sound speed $\sim 1$.} In particular, the $ k^4$-term in \eqref{eq:tgi} is  subdominant for all the modes captured by the effective description. Then, the system undergoes a transient unstable phase triggered by  $c_s^2(\tau)=-\tilde{c}_s^2<0$ for $\tau_t<\tau<\tau'_t$, and the  amplitude of (some of) the modes  increases exponentially. In the last phase starting at $\tau'_t$, the original value of the sound speed is restored and the modes stop growing. To understand the effect of the instability on the power spectrum, it is convenient to follow the evolution of modes {with different momenta} $k$. Let us call $\tau_k$ the conformal time at which a certain mode with momentum $k$ exits the horizon. Modes with sufficiently small $k$, i.e.~$\vert k \tau_t \vert\ll1$, that exit the horizon before the unstable phase, i.e.~$\tau_k<\tau_t$, do not feel the effect of the gradient instability and  will not have an enhanced  power spectrum.
Let us then focus on modes that exit the horizon during or after the unstable phase, i.e.~such that  $\tau_k>\tau_t$. These will in general grow because of the instability and will have a 
power spectrum enhanced (with respect to the previous ones) by a factor of $\sim\e^{2\tilde{c}_sk\Delta\tau}$, where $\Delta\tau=\tau_k-\tau_t$ or $\Delta\tau=\tau'_t-\tau_t$, whichever is the smallest. This seems to indicate that the modes with the largest enhancement are the ones with the largest momentum. However, this comes with a caveat.  One should take also into account the presence of the $ k^4$-term in \eqref{eq:tgi}, which can be relevant in the unstable phase since $\tilde{c}_s\ll1$. Indeed, modes with large enough $k$, in particular those with $\frac{\tilde{c}_s}{\sqrt{\alpha}}\lesssim \vert k\tau'_t \vert  < \vert k\tau_t \vert \lesssim\frac{1}{\sqrt{\alpha}} $, whose dynamics resembles the one of ghost inflation, will not be significantly affected by the instability and will not grow as much as those with intermediate $k$.
Thus, the largest relative enhancement at the level of the power spectrum is obtained for modes with $\frac{\tilde{c}_s}{\sqrt{\alpha}} \gtrsim \vert k\tau_t \vert   $, and corresponds to $\sim\e^{\frac{2\tilde{c}_s^2}{\sqrt{\alpha}} (1-\tau'_t/\tau_t)}$, where we assumed $\vert\tau_k\vert<\vert\tau'_t\vert$.\footnote{Note that this estimate reproduces the exponential factor in \eqref{psenh-2} in the limit $\vert\tau'_t/\tau_t\vert\ll1$.}
An explicit example is shown in Figure~\ref{fig:TransitionPNP-L}, where we plot the 
power spectrum $\Delta_\zeta^2$ as a function of $k$, normalized  to the CMB value, $ 2 \times 10^{-9}$, at $k\ll k_\text{peak}$.\footnote{Assuming the scaling \eqref{M2Mb2h} for $\bar{M}_2^2$  (see also Table \ref{tab1} in Appendix \ref{app:sizeoperators}), with the chosen value $\alpha=10^{-4}$ for $\alpha$ defined in \eqref{cs}, to have the power spectrum in Figure~\ref{fig:TransitionPNP-L}   correctly normalized at $k/k_\text{peak}\ll1$ to the CMB value, one needs $\alpha_1 \sim \mathcal{O}( 10-100)$.} The plot has been obtained by solving the equation \eqref{eqsrst} with $c_s^2=\bar{c}_s^2$ for $\tau<\tau_t$ and $\tau>\tau'_t$, and the equation \eqref{eq:tgi} with $c_s^2=\tilde{c}_s^2$ for $\tau_t<\tau<\tau'_t$, and then matching $\zeta_k$ and $\zeta'_k$ across the transition points.
The shape with a peak at intermediate $k$ is in agreement with the numerical solution of Section~\ref{section:TransitionPosNegPos}. Note that the oscillations in Figure~\ref{fig:TransitionPNP-L} follow from interference effects resulting from the transition between different phases \cite{Ballesteros:2018wlw}.\footnote{We stress that this is true for all the oscillations visible in Figure~\ref{fig:TransitionPNP-L}, except those at the rightmost part of the plot. These ones are instead {affected by}
the fact $\e^{-i \bar{c}_s k \tau_i}$ (with $\tau_i$  being the initial time where the boundary condition is imposed)  is no longer the correct initial condition for  modes with very large $k$.} 

\subsection{Numerical solution}\label{section:TransitionPosNegPos}

\begin{figure}[h!] 
\begin{center} 
\includegraphics[scale=0.6,trim=3.1cm 13cm 1cm 4cm,clip]{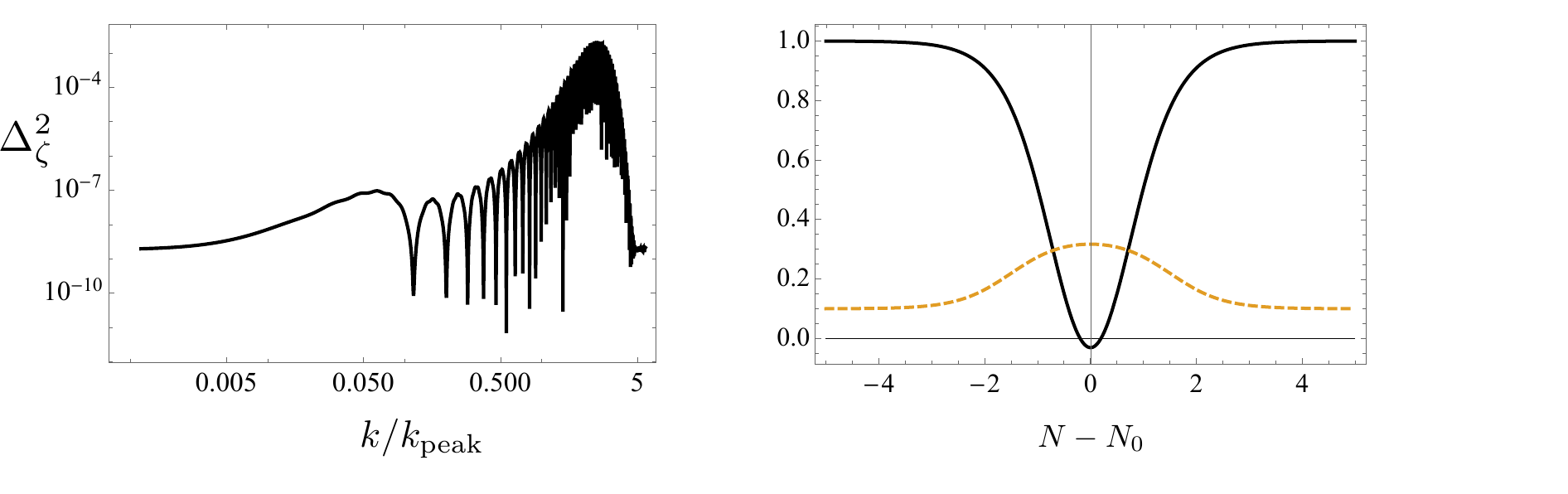}
\caption{
On the left we plot the power spectrum obtained by solving  \eqref{eq:tgi} numerically,  with the parameters shown on the right. The yellow dashed line is $10^6\alpha$ and the black line is $c_s^2$. The (e-fold) time dependence of $c_s$ is given by $ c_s^2=(1-(\tanh(N-N_0+0.5)-\tanh(N-N_0-0.5))\delta_{c_s})c_{0s}^2$ with $c_{0s}^2=1$, and $\delta_{c_s}=1.115$ a, whereas $\alpha=(\tanh(N-N_0+1.5)-\tanh(N-N_0-1.5))\delta_{\alpha}+\alpha_0$ with $\alpha_0=10^{-7}$, and $\delta_\alpha=1.2\times 10^{-7}$. We assume that all the other parameters are  constant. In particular, we set  $\varepsilon=10^{-4}$ and $\alpha_1=150$, while we set all  the other slow-roll parameters  to zero. }
\label{fig:TransitionSRGISR}
\end{center} 
\end{figure}

In this section, we present a numerical solution to the equation \eqref{app:MSeq2} for a smoother version of the transition described in the previous section. The result is reported in Figure~\ref{fig:TransitionSRGISR}.
To obtain the solution, we choose $-0.03 \lesssim c_s^2 \lesssim1$, with $c_s^2 \simeq1$ before and after the transition,   while $\alpha\sim \mathcal{O}(1) \times 10^{-7}$ throughout the entire evolution. 
We  assume  that the system starts in a phase of slow-roll evolution and that the dynamics of the perturbations is initially governed by a standard dispersion relation with $c_s^2\simeq 1$, as displayed in the right panel of Figure~\ref{fig:TransitionSRGISR} (solid line). Then, the sound speed gradually diminishes
and, after  a couple of e-folds, the system enters the ghost-condensate-like phase with  
quadratic dispersion relation. The dashed curve in  the right plot of Figure~\ref{fig:TransitionSRGISR} describes the evolution of {the EFT coefficient $\alpha$}.
Thereafter, the system temporarily stays in a phase with $c_s^2<0$. This triggers an exponential growth of the amplitude of the perturbations, resulting in an enhanced power spectrum as shown in the left plot of Figure~\ref{fig:TransitionSRGISR}.  Afterwards, $c_s^2$ returns positive restoring the original dynamics.

{
In the plot in Figure~\ref{fig:TransitionSRGISR}, one recognizes the two salient  features discussed in Sections~\ref{sec:gi-to-ics} and \ref{app:rsexpenh}: the first one is the exponential enhancement obtained during the unstable phase {(as opposed to the example in Figure~\ref{fig:SRtoGI} where the growth is compatible with a power law)};  the second one is the sudden decay in the shape of the power spectrum  after the peak.
The presence of the  {plateau} before  the exponential growth simply follows from the fact that modes with sufficiently small $k$, which exit the horizon very soon, remain unaffected by the instability. The second plateau, occurring after the peak {(and barely shown in the plot)}, arises instead, as noticed already in Section~\ref{app:rsexpenh}, because modes with very large $k$ remain  dominated by the quadratic dispersion relation during the transient phase with $c_s^2<0$. Therefore, they effectively remain in the ghost-condensate-like regime and are  also not enhanced.
As a result, the final power spectrum is peaked only over a finite range of momenta,  which are those that effectively experience the gradient instability.
}

{
At this point, one might worry about the consistency of the effective expansion in the unstable phase and, in particular, across the points where the dispersion relation for  $\omega^2$ turns from positive to negative, and {vice versa}. A prototypical example of the  instability is the (an)harmonic oscillator in quantum mechanics with inverted  potential. Let us imagine  that the curvature of the  potential is  initially positive 
and that the oscillator is 
in its ground state. At some time,
the curvature of the potential suddenly turns 
negative. In the limit of an instantaneous transition, the system will initially still be in the same state, but the wavefunction will start to change in time: 
it will grow and spread  as a result of the unbounded potential.
However, as long as the unstable phase is just temporary---one can imagine that, after a certain time interval, the curvature of the  potential 
becomes positive again---{the wave function will stabilize.}
In the case of an EFT with transient gradient instability, one should check that the maximum rate of the instability lies within the realm of validity of the effective expansion. In particular, focusing on the dispersion relation in the unstable phase, which we can write schematically as 
\begin{equation}
\omega^2 H^{-2} \sim -\vert c_s\vert^2 (k \tau)^2 + \alpha (k\tau)^4 \, ,
\end{equation}
the maximum rate of the instability is given by
\begin{equation}
\Gamma \equiv \text{max}(\text{Im }\omega) \sim \frac{\vert c_s\vert^2H}{\sqrt{\alpha}} \sim \vert c_s\vert^2\Lambda_3 \, ,
\label{gammagi}
\end{equation}
where we used \eqref{alphaL3}. 
{Taking now into account Eq.~\eqref{scs-ghost}} and  that $\vert c_s\vert^2 \ll 1$ in the unstable phase,
from \eqref{gammagi} one finds that $\Gamma \ll \Lambda_\star$, corresponding therefore to energies that are below the cutoff.\footnote{This is correct if the loop corrections are dominated by the high-energy part of the loop integral, i.e.\ by the  dynamics of modes with quadratic dispersion relation, $\omega\propto k^2$, which are perfectly stable and weakly coupled. However, it does not completely address  the question of strong coupling in the unstable phase. A quick scaling argument, following the one in the next section, seems to suggest the presence of potentially dangerous large  $N$-loop corrections in the large-$N$  limit, where the leading contribution comes from the exchange of exponentially growing modes.  We leave a more detailed analysis of this aspect for future work.}
Finally, another possible obstacle in applying the EFT to describe the transition 
between slow-roll inflation (or the ghost condensate) and the unstable regime is that a too fast 
evolution may jeopardize the estimate of the strong coupling scale  across 
phases---see 
our discussion in Section~\ref{subsec:strongcoupling} above. On the other hand, this obstacle {may}
not {be} insurmountable as long as one does not need to resolve the details of the transition between the two phases. 
Since we are {mainly} interested in 
the computation of the global shape of the power spectrum, we can still use the EFT to describe the relevant phases, whose robustness is guaranteed by symmetries, and match across the transition points, as we already schematically did before  in the previous section.
}

\section{Non-Gaussianities and the validity of perturbation theory}
\label{sec:ng-pt}

The main reason that   has recently  motivated  the study of inflationary scenarios  with large power spectra on small scales is the possibility of generating PBHs. {Analytical and numerical analyses indicate that $\zeta\gtrsim 0.1$ is needed for PBH formation.}
Therefore, in order to correctly estimate  their abundance, the  information about the size of the two-point function {may not be} enough, and one {may need to} know additional information about the full shape of the distribution, {e.g.\  the form of} higher-{order} correlation functions. The most promising approach {to this issue may lie on techniques that attempt to go} beyond perturbation theory, see e.g.\ \cite{Panagopoulos:2019ail, Ezquiaga:2019ftu, Figueroa:2020jkf, Pattison:2021oen, Celoria:2021vjw, Biagetti:2021eep}  for several recent attempts with different methods.
Our goal in the present section is 
more modest. We will estimate the size of the leading $n$-point correlation function for the example presented in the previous section and  compare it with the power spectrum via the relation
\begin{equation}
\frac{\langle\zeta^n\rangle}{\langle\zeta^2\rangle^{n/2}}\lesssim1 \,,
\label{validitypt}
\end{equation}
which may give an indication of the validity of the perturbative expansion. 

The presence of an (even temporary) instability inducing an exponential growth of the curvature perturbations may be worrisome, as it may yield  even larger effects at the level of higher-point correlators, indicating the breakdown of perturbation theory. The actual situation is, however, less catastrophic. Refs.~\cite{Bjorkmo:2019qno,Fumagalli:2019noh} have shown indeed that, during a gradient instability phase, higher-point correlation functions of $\zeta$  grow less than the naive counting based on the scaling of the field would suggest. In the present section, we  briefly summarize this result and extend it to the case of the interactions resulting from the operators $(\delta g^{00})^n\delta K$ in the effective theory \eqref{eftofi}. 

A generic $n$-point correlation function can be computed  in the in-in formalism \cite{Weinberg:2005vy} as
\begin{equation}
\langle \zeta^n(t) \rangle = \sum_{N=0}^\infty i^N \int_{-\infty}^t \D t_N \ldots \int_{-\infty}^{t_2} \D t_1\langle
\left[ H_I(t_1), \cdots \left[  H_I(t_N), \zeta^n_I(t) \right]
\right]
\rangle \, ,
\label{Winin}
\end{equation}
where $H_I$ represents an insertion of the interaction Hamiltonian. After the Wick contractions of the fields, each nested commutator corresponds 
to the imaginary part of the product of a certain number of  mode functions $\zeta_k$ with an equal  number of complex conjugate modes $\zeta^*_k$. Since, from \eqref{eq:GIcurvature}, \eqref{A} and \eqref{B}, the mode function for the curvature perturbation in the unstable phase is schematically of the form
\begin{equation}
\zeta_k =  \e^{i \psi} \mathcal{N} \left[\e^{\frac{c_s^2}{\sqrt{\alpha}}+c_s k \tau}(1-c_s k \tau)  - i \, \e^{-\frac{c_s^2}{\sqrt{\alpha}}-c_s k \tau}(1+ c_s k \tau)  \right] \, ,
\label{zetaschem}
\end{equation}
where $\psi \in \mathbb{R}$ is some  phase and $\mathcal{N} $ an irrelevant, constant, (and also real) normalization factor, it is clear that, when taking the imaginary part of the product of an $m$-number of $\zeta_k$ and an $m$-number of $\zeta_k^*$, the leading contribution, in the limit  $x\equiv c_s^2/\sqrt{\alpha}\gg1$, is  $\e^{(2m-2)x}$, and not $\e^{2mx}$. The insertion of at least one  decaying mode in \eqref{zetaschem} is necessary to obtain something that is imaginary and yields a non-zero commutator. In practice, this means that, at tree-level, given a generic correlator \eqref{Winin},  the leading exponential scaling in  $x$ can be obtained as follows: one should attach a factor of $\e^{2x}$ to each (internal or external) leg in the diagram and a factor $\e^{-2x}$ to each vertex (which corresponds to a single insertion of the interaction Hamiltonian)  \cite{Bjorkmo:2019qno}. For instance, with $(n-2)$ insertions of the cubic Hamiltonian, the $n$-point correlator scales therefore at most as $\langle\zeta^n\rangle \sim \e^{(2n-2)x}$ \cite{Bjorkmo:2019qno}. In addition to the exponential enhancement, as pointed out in \cite{Fumagalli:2019noh}, there could also be a polynomial  in $x$ multiplying the  exponential factor. These additional contributions can  be understood as follows. To be concrete, let us focus for simplicity on  our example of Section~\ref{sec:gi-to-ics}. When applying the formula \eqref{Winin}, one needs to perform a certain number of integrations in time. Since we are interested in estimating the leading contribution to the correlator coming from the instability,  we can simply restrict, for this purpose,   the integration interval over the unstable phase, which is defined to start at $\tau=\tau_0$, and cut the integrals from below at $\tau_0$.\footnote{We are essentially disregarding terms that correspond to integrals of oscillating functions, which are not subject to the exponential enhancement. They provide, therefore, subleading corrections to the correlator in the large-$x$ limit.} 
From \eqref{zetaschem}, one infers that $\zeta$   scales in conformal time as $\zeta\sim \e^{c_s k \tau} k \tau$.\footnote{Since we will eventually replace $\tau$ with $\tau_0$, we are simply taking the large-$\vert\tau\vert$ limit in \eqref{zetaschem}.} Thus, the interaction Hamiltonians that provide the largest scaling in $\tau$ are those with the largest number of temporal ($\partial_t = -H \tau \partial_\tau$) or spatial ($\frac{\partial_i}{a}= -H \tau \partial_i$) derivatives per number of fields. For the EFT \eqref{eftofi},  these Hamiltonians arise from the higher derivative operators $(\delta g^{00})^n\delta K$ and  are schematically of the form  $H_I^{(m)}= \int \D^3\bfx \, a^3 (\tau \partial_\tau \zeta)^{m-1} \tau^2 \partial_i^2\zeta $ \cite{Pirtskhalava:2015zwa,Pirtskhalava:2015ebk},\footnote{The derivatives acting on $\zeta$ in $H_I^{(m)}$ could  in principle be either in  conformal time or in the  spatial coordinates. The final counting does not change.} which thus  scale roughly as  $\sim \tau^{2m-2}$. Then, plugging into  the formula \eqref{Winin}, in addition to the exponential  factor, the correlator takes schematically the form
\begin{equation}
\langle \zeta^n(\tau) \rangle \supseteq \sum_{N} \e^{\left( \sum_{j=1}^N m_j - 2 N +n \right)x}  \int_{\tau_0}^\tau \D \tau_N \, \tau_N^{2m_{_{N}}-3} \ldots \int_{\tau_0}^{\tau} \D \tau_1 \,  \tau_1^{2m_1-3} 
 \, ,
\label{scalenpf}
\end{equation}
where we dropped all the dependence on the spatial momenta and  retained only the leading polynomial scaling in $\tau$ in each integral.
The polynomials in $\tau$ in \eqref{scalenpf} are usually harmless because they multiply an exponential in $\tau$ and are converted by the integral into  functions of the external momenta. However, we mentioned above that each vertex involves a decaying mode. Thus, the situation changes if the spatial momenta are such  that the momentum of the decaying mode exactly equals the sum of the moduli of the momenta of the growing modes \cite{Bjorkmo:2019qno,Fumagalli:2019noh}. In such a flattened configuration, the exponential  factor in the integral drops and one is left with just an integral of the type  $\int \D \tau_j \tau_j^{2m_j-3} $, which yields $\tau_0^{2m_j-2}$.\footnote{We come back to this point in Appendix~\ref{app:ng-pt} where we present an explicit calculation of the three-point correlation function.}
Replacing $\tau_0$ with $\sim c_s /(k\sqrt{\alpha})=  x /(kc_s)$, Eqs.~\eqref{scalenpf} and \eqref{psenh1} then yield
\begin{equation}
\frac{\langle \zeta^n \rangle}{\langle \zeta^2 \rangle^{n-1}} \sim \e^{-2x(n-1)}\sum_{N} \e^{\left( \sum_{j=1}^N m_j - 2 N +n \right)x}  x^{2\sum_{j=1}^N m_j-2 N} 
= \sum_{N}  x^{2n + 2N -4}
 \, ,
\label{scalenpf-2}
\end{equation}
where we used the relation $n= \sum_{j=1}^N m_j - 2 N + 2$, which is valid at tree level, and where we assumed for simplicity $c_s\sim \mathcal{O}(1)$.\footnote{Taking $\vert c_s\vert \ll 1 $ would simply make the ratio \eqref{scalenpf-2} between higher-point correlators and the power spectrum larger---see Appendix~\ref{app:ng-pt} for further details.}
The largest enhancement is obtained when the number of vertices in the diagram is the highest, i.e.\ when $N=n-2$, corresponding to $\langle \zeta^n \rangle/ \langle \zeta^2 \rangle^{n-1} \sim x^{4n -8}$.\footnote{{Note that this result is different from the one of  \cite{Fumagalli:2019noh} (which is $\langle \zeta^n \rangle/ \langle \zeta^2 \rangle^{n-1} \sim x^{3n -6}$), where the couplings of  the operators $(\delta g^{00})^n\delta K$  are set to zero.}}
In the large-$n$ limit, Eq.~\eqref{validitypt} reduces to the condition $\sqrt{\Delta_\zeta^2x^8} \lesssim1$. Using \eqref{psenh-2}, this condition tells us that  perturbation theory  might not be completely under control when $x\gtrsim 6$. This bounds the maximum power spectrum reachable in the example of the previous section to be at most $\Delta_\zeta^2 \lesssim 10^{-5}$.\footnote{{It is easy to show that the bound derived from requiring that the energy density associated with the exponentially growing modes remains small compared with the background energy density---which is a necessary condition to avoid backreaction, see e.g.~\cite{Holman:2007na}---is less strong than the bound obtained here.}}
This bound should be taken with a grain of salt. Even though perturbative unitarity is satisfied thanks to the modified dispersion relation with the $k^4$-term that dominates at high momenta in the unstable phase ({provided that $H<\Lambda_*$,} see the considerations in Section~\ref{subsec:strongcoupling}), a stronger constraint may result from loop corrections to the power spectrum computed using \eqref{Winin}. This would require to extend the considerations in the present section beyond tree level, which we leave for future work. 
 In Appendix~\ref{app:ng-pt}, we present an explicit calculation confirming the scaling \eqref{scalenpf-2}, where we will also track down the scaling in the sound speed, which  is relevant when $\vert c_s\vert\ll 1$.

\section{Discussion}
\label{sec:conclude}

Motivated by the idea that  PBHs---produced from the collapse of overdensities originating from quantum inflationary fluctuations---could provide an explanation for dark matter, we have studied to what extent it is possible to obtain a large power spectrum for the curvature perturbation within the effective theory of single-field inflation. A large power spectrum (on distance scales smaller than the CMB ones) is a necessary ingredient to induce those overdensities. A correct estimate of the  abundance of PBHs requires good knowledge of the shape of the tail of the probability distribution of the curvature fluctuation and {it might be that, in general, it can only be tackled accurately} with non-perturbative methods. From this perspective, our goal in this paper has been more modest. However, understanding which dynamics can lead to a large power spectrum is a prerequisite for a complete picture of PBH formation in the EFT of inflation. Using this {rather} model-independent framework {(which only assumes a quasi-de Sitter universe and the existence of a single degree of freedom that spontaneously breaks  time diffeormorphisms)} we have analyzed several mechanisms that, by changing the dynamics of the perturbations during inflation,  induce  an enhancement in the 
power spectrum. 
{One of our main results has been showing that} a transition into a ghost-inflation-like phase with quadratic dispersion relation  can  increase the power spectrum by several orders of magnitude and, at the same time, alleviate the strong coupling problem arising in models with small sound speed.  The enhancement  scales  as $(\Lambda_3/H)^{3/2}$, where $\Lambda_3$ is the energy scale associated with higher derivative operators in the underlying scalar-tensor theory, although the actual amplitude of the peak in the power spectrum depends on the duration of the ghost-inflation-like phase and   the details of the transition, which may result in additional oscillations (see Figure~\ref{fig:SRtoGI}). 
Assuming the scaling $(\Lambda_3/H)^{3/2}$, a power spectrum of order of $\Delta_\zeta^2\sim10^{-2}$ would require $\Lambda_3\sim 3 \times 10^{4} H$, which may be obtained at the price of further suppressing higher derivative operators  with respect to more standard $P(\phi,(\partial_\mu \phi)^2)$ operators in the underlying covariant theory.  An explicit realization and model building in this context is left for future work.
In addition, we have considered the case where an exponentially growing mode is temporarily turned on, as a result of a transient gradient instability in the spectrum of the perturbations.
{At any fixed time during this transient phase,  only the behavior of the low-energy modes is affected by the instability, while the dynamics of the perturbations at large physical momenta is under control thanks to a modified dispersion relation. 
} 
The resulting enhancement in the power spectrum goes in this case as $\e^{2\vert c_s\vert^2 \Lambda_3/H}$, where $\vert c_s\vert$ is the absolute value of the sound speed in the unstable phase. 
This estimate was obtained assuming instantaneous transitions and constant sound speed in the different phases.   
Here, a milder separation between $\Lambda_3$ and $H$ is enough to reach large values for the power spectrum.

There are several important or interesting aspects  that we have left for future research.
\begin{itemize} 

\item[$\ast$] In addressing the question on the validity of  perturbation theory  in the scenario with the  gradient instability, we have used simple tree-level scaling arguments (which are a straightforward  generalization of similar analyses done in \cite{Bjorkmo:2019qno,Fumagalli:2019noh}) to infer the behavior of a generic $n$-point correlator.
For this purpose, we have not  taken into account possible combinatorial factors that might be relevant in the large $n$ limit.  It would be interesting to understand in detail the seeming presence of a factorial enhancement in the cosmological $n$-point correlators {in the scenarios we have studied as well as in  generic} single-field models of inflation, as done in {a concrete} multi-field context in \cite{Panagopoulos:2020sxp}.

\item[$\ast$]
We have not fully addressed the issue of strong coupling in the gradient instability phase. Loop corrections that are dominated by the high-energy part of the loops are expected to remain under perturbative control, since the dynamics at large $k$ is dominated by the $k^4$-term, which is perfectly stable and weakly coupled. On the other hand, one may be worried that $N$-loop corrections to tree-level amplitudes may soon become dominated by the exchange of unstable modes,  and  grow dangerously as $\sim \e^{2Nx}$, in the large-$N$ limit, as a  naive scaling reasoning suggests. It would be interesting to check if this is indeed the case by performing a more detailed  calculation of loop corrections  to the power spectrum (see, e.g.\ \cite{Senatore:2009cf,Melville:2021lst}) in the cases we have explored. 

\item[$\ast$] A detailed calculation of the non-Gaussian properties of $\zeta$ as well as an investigation into its stochastic dynamics in the scenarios we have delineated may help to characterize better the PBH abundance and determine to which extent the power spectrum needs to be large for PBHs to account for all dark matter. These pending analyses are doubly important in connection with the improved perturbative bounds mentioned above. 

\item[$\ast$] A significant stochastic background of gravitational waves is generically induced at second order in perturbations if the scalar power spectrum is large. The location of the peak of the latter determines the peak frequency of the induced gravitational wave background (see e.g.~\cite{Ballesteros:2020qam}) and features such as oscillations can be carried over from one spectrum to the other, see \cite{Fumagalli:2020nvq}.  A characterization of the correlations between the two may help to distinguish our scenarios from others that are also motivated from PBH formation, such as e.g.\ \cite{Ballesteros:2020qam, Fumagalli:2020nvq, Braglia:2020taf}.

\item[$\ast$] Finally, a gradient instability often appears as well in bouncing cosmologies (see, e.g., \cite{Libanov:2016kfc,Kobayashi:2016xpl,Pirtskhalava:2014esa,Creminelli:2016zwa,Cai:2016thi}). It might be interesting to understand whether  such an instability can also lead in those scenarios to a large power spectrum.
This would open   to addressing the formation of PBHs in such alternative cosmologies, like bouncing and genesis models.\footnote{See \cite{Chen:2016kjx} for previous works on PBH formation in bouncing cosmological scenarios.}

\end{itemize}

\paragraph{Acknowledgements.} 
We thank Jose Beltr\'an Jim\'enez, {Jacopo Fumagalli,} Sebastian Garcia-Saenz, Lam Hui, Alberto Nicolis, Gonzalo Palma, Mauro Pieroni, S\'ebastien Renaux-Petel, Enrico Trincherini {and Lukas Witkowski} for useful discussions and comments.
The work of GB and SC has been funded by a Contrato de Atracci\'on de Talento (Modalidad 1) de la Comunidad de Madrid (Spain), number 2017-T1/TIC-5520 and the IFT Centro de Excelencia Severo Ochoa Grant SEV-2016-0597. The work of GB has also been funded by MCIU (Spain) through contract PGC2018-096646-A-I00. LS is supported by Simons Foundation Award No.~555117.

\appendix

\section{Estimating the size of the effective couplings}
\label{app:sizeoperators}

Our starting point was the EFT of inflation \eqref{eftofi} \cite{Creminelli:2006xe,Cheung:2007st}. As any other EFT, the  action  \eqref{eftofi} is organized as  an infinite sum of higher dimensional operators, and  is completely specified by the low-energy degrees of freedom (which fix the form of the operators in the Lagrangian)\footnote{In the present context the low-energy degrees of freedom are the  scalar  and the two graviton degrees of freedom, even though, in the main text, we primarily worked in the decoupling limit \cite{Cheung:2007st}, which allowed us to neglect the coupling to gravity.} and the symmetries at play in the system (which dictate a set of rules to power count the size of the operators). In the absence of  symmetries, higher derivative operators like $ \hat{M}_1^3 \delta g^{00}\delta K$ in \eqref{eftofi} are expected to be  subleading at low energies. 
In other words, they can never provide $\mathcal{O}(1)$ corrections at the level of the observables without incurring in fine-tuning problems or having infinitely many operators becoming of the same size at low energies,  invalidating therefore the predictive power of the theory.
This conclusion can change if the underlying scalar-tensor theory has a weakly broken galileon (WBG) invariance \cite{Pirtskhalava:2015nla}. 
The approximate symmetry introduces  a different power counting rule for some of the operators in the theory, allowing their couplings to be larger than what they would normally be without the symmetry. In the context of \eqref{eftofi}, this  allows to have a well-defined hierarchy $M_2^4 \sim  \hat{M}^3_1H$ in the theory. For details, see \cite{Pirtskhalava:2015nla,Santoni:2018rrx} (and also \cite{Goon:2016ihr}). We will   briefly summarize now  the main ingredients and results, and report in Table~\ref{tab1} the  ranges of values for the effective parameters in  \eqref{eftofi} that are allowed by the (weakly broken) symmetry.

Let us start considering the most general  theory of a scalar field coupled to gravity. Let $\Lambda_3$ be the cutoff of the EFT and let us introduce the scale $\Lambda_2 = (\Mpl\Lambda_3^3)^{1/4} \gg\Lambda_3$, whose relation to $\Lambda_3$ is going to be fixed by the coupling to gravity. The WBG symmetry allows to classify the operators in the EFT schematically as follows \cite{Pirtskhalava:2015nla,Santoni:2018rrx}:
\begin{equation}
L \supseteq
c^{\rm I}\underbrace{\frac{(\nabla\phi)^{2n}}{\Lambda_2^{4(n-1)}}\frac{(\nabla\nabla\phi)^m}{\Lambda_3^{3m}}   }_\text{(I)}
+c^{\rm II} \underbrace{\frac{(\nabla\phi)^{2n}}{\Lambda_2^{4n}}\frac{(\nabla\nabla\phi)^m}{\Lambda_3^{3m-4}}  }_\text{(II)}
+ c^{\rm III}\underbrace{\frac{\nabla^m(\nabla\nabla\phi)^n}{\Lambda_3^{3n+m-4}}  }_\text{(III)}
 \, ,
 \label{eftrob}
\end{equation}
where: (I) refers to operators which benefit from  non-renormalization properties \cite{Nicolis:2008in,Luty:2003vm,Pirtskhalava:2015nla,Santoni:2018rrx,Goon:2016ihr,Noller:2019chl}, whose couplings $c^\text{I}$ are protected against large quantum corrections (in particular, $\delta c^\text{I}/c^\text{I}\sim \Lambda_3/\MP$ \cite{Pirtskhalava:2015nla,Santoni:2018rrx}); (II) represents  operators with the same number of fields and derivatives but that are not protected by the symmetry, i.e.\ $\delta c^\text{II}/c^\text{II}\sim\mathcal{O}(1)$; (III) contains all the other  operators with at least two derivatives per field that are  generated at the scale $\Lambda_3$ and that are such that $\delta c^\text{III}/c^\text{III}\sim\mathcal{O}(1)$ \cite{Santoni:2018rrx}.\footnote{Note that these operators are usually associated with ghost degrees of freedom. However, since they enter at the cutoff scale, they are completely harmless at low energies.}
The WBG symmetry guarantees that  the parametric separation  between $\Lambda_2$ and $\Lambda_3$, and the power countings in \eqref{eftrob}  are stable against quantum corrections. For instance, let us consider the following explicit example,
\begin{equation}
L =  \Lambda_2^4 P(X) + (\nabla\phi)^2\frac{\square\phi}{\Lambda_3^3}
 + \frac{(\square\phi)^2}{\Lambda_3^2}    \, ,
 \qquad
X\equiv -\frac{(\nabla\phi)^2}{\Lambda_2^4}\, .
\label{eftrob-2}
\end{equation}
Schematically, to rewrite \eqref{eftrob-2} in the  language of the action \eqref{eftofi} in the unitary gauge, we can replace $\square \phi \sim \sqrt{X} K$ (see, e.g., \cite{Gleyzes:2013ooa}). The $P(X)$ operators  simply correspond to the terms $(\delta g^{00})^n$ in \eqref{eftofi}, while $(\nabla\phi)^2 \square\phi$  \cite{Deffayet:2010qz} contributes to the operator $\delta g^{00}\delta K$ \cite{Creminelli:2006xe,Cheung:2007st}. The last term in \eqref{eftrob-2} schematically becomes instead  $(\delta K)^2$ in \eqref{eftofi}.
The typical sizes of the effective coefficients  in \eqref{eftofi} in the limiting cases of a potentially-driven evolution (characterized by  $X\sim \sqrt{\varepsilon}\ll1$ on the background)  and  a kinetically-driven evolution (which has instead $X\sim 1$  on the background) have been discussed in \cite{Pirtskhalava:2015nla,Pirtskhalava:2015zwa, Pirtskhalava:2015ebk,Santoni:2018rrx}  and are summarized  in Table~\ref{tab1} below.
\renewcommand{\arraystretch}{2.2}
\begin{table}[h]
\centering
\begin{tabular}{|c||c|c|c|c|c|c|c|cl}
  \hline
Potentially-driven evolution & Kinetically-driven evolution  \\
 \hline\hline
 $ \Lambda_2^4\sim \sqrt{\varepsilon} \MP^2H^2 $, \qquad $ \Lambda_3^3\sim \sqrt{\varepsilon} \MP H^2 $ & $ \Lambda_2^4\sim  \MP^2H^2 $, \qquad $ \Lambda_3^3\sim  \MP H^2 $  \\
$ M_2^4\sim \varepsilon \MP^2H^2 $, \qquad
$ \hat M_1^3\sim \varepsilon \MP^2H $, 
& $ M_2^4\sim  \MP^2H^2 $, \qquad
$ \hat M_1^3\sim  \MP^2H $, \\
$ \bar M_2^2\sim \sqrt{ \varepsilon} \MP\Lambda_3 \sim \varepsilon^{2/3} \MP^{4/3}H^{2/3} $ &  $ \bar M_2^2 \sim \MP\Lambda_3  \sim  \MP^{4/3}H^{2/3} $ \\
  \hline
\end{tabular}\\
\caption{Typical sizes of the scales $\Lambda_2$ and $\Lambda_3$, and the effective couplings in \eqref{eftofi}. The left column reports the typical sizes for these quantities when computed around background solutions that are mostly dominated by the potential (with $X\sim \sqrt{\varepsilon}$) \cite{Pirtskhalava:2015nla,Pirtskhalava:2015zwa}, while the higher derivative  operators provide $\mathcal{O}(1)$ corrections to the dynamics of the perturbations. The right column  shows instead the values when the background dynamics is mainly driven by the derivative operators (i.e., $X\sim 1$) \cite{Pirtskhalava:2015nla, Pirtskhalava:2015ebk}.   In both cases, $\square\phi\sim\Lambda_3^3$ on the background.
The typical sizes of $ M_2^4$ and $ \hat M_1^3$ are computed in \cite{Pirtskhalava:2015nla,Santoni:2018rrx}. The size of  $\bar M_2^2$ can be estimated by using that
$\square\phi  \sim \Lambda_2^2\sqrt{X}  K$ (see, e.g., \cite{Gleyzes:2013ooa}) and therefore $(\square\phi)^2/\Lambda_3^2\sim (\Lambda_2^4 X/\Lambda_3^2) K^2$, which implies $\bar{M}_2^2 \sim \Lambda_2^4 X/\Lambda_3^2 $.  }
\label{tab1}
\end{table}

\section{Non-Gaussianities and the validity of perturbation theory: an explicit example}
\label{app:ng-pt}

In Section~\ref{sec:ng-pt} we estimated the size of non-Gaussianities and the breakdown of perturbation theory in the EFT of single-field inflation in the presence of higher derivative operators if the curvature perturbations undergo a transient phase of gradient instability, extending the arguments of \cite{Bjorkmo:2019qno,Fumagalli:2019noh}. In this section, we want to consider an explicit example and show that the correlators satisfy indeed the scaling discussed in  Section~\ref{sec:ng-pt}. For simplicity, we assume that the cosmic  evolution is first well-described by a stage of slow-roll with $c_s^2>0$. Then, at some conformal time $\tau=\tau_0$, the system enters an unstable phase with $c_s^2<0$. In order to provide analytic results and simplify the final expressions, we will assume that the transition is instantaneous and we will implement it just by switching  $c_s\rightarrow i c_s$, while keeping the absolute value of $c_s$ unchanged. Even if this is  an extreme approximation  of the more feasible scenario that we presented and solved numerically  in Section~\ref{section:TransitionPosNegPos}, this example contains all the ingredients that we need to illustrate more concretely the considerations outlined in Section~\ref{sec:ng-pt}.

The solutions to \eqref{app:MSeq2} (with $\alpha=0$) in this stepwise approximation are given by
\begin{align}
\label{app:mode1}
v_k(\tau)&  = \frac{\e^{-i c_s k \tau }}{\sqrt{2 c_s k}} \left(1-\frac{i}{c_s k \tau }\right)  \, ,
\qquad
\text{for } \tau<\tau_0 \, ,
\\
\label{app:mode2}
v_k(\tau) & = A_k \e^{c_s k \tau } \left(1-\frac{1}{c_s k \tau }\right)+B_k \e^{-c_s k \tau } \left(\frac{1}{c_s k \tau }+1\right)  \, ,
\qquad
\text{for } \tau>\tau_0 \, ,
\end{align}
where we imposed  Bunch--Davies  initial conditions, and where $A_k$ and $B_k$ are determined by requiring continuity of $\zeta_k$ and $\zeta'_k$ across $\tau_0$,
\begin{align}
A_k & = (-1)^{-1/4}\frac{  (c_s k \tau_0 +1-i) \e^{  - i c_s k \tau_0} }{2(c_s k)^{3/2} \tau_0} \e^{- c_s k \tau_0  } \, ,
\\
B_k & = (-1)^{1/4}\frac{  (c_s k \tau_0 -1-i) \e^{  - i c_s k \tau_0} }{2(c_s k)^{3/2} \tau_0} \e^{ c_s k \tau_0  } \, .
\end{align}
As discussed in Section~\ref{sec:ng-pt}, the operators that   yield the largest contribution at the level of the correlators, with  the strongest dependence on the duration of the unstable phase, are those with the largest number of derivatives per field. In the small-$\vert c_s\vert$ limit, these are of the form $(\partial_i\zeta)^{2n}\partial^2\zeta$, with positive $n$.
As an illustrative example, let us compute the three-point correlation function   with the  interaction Hamiltonian $H_I^{(3)}= \int \D^3 \bfx \, a^{-1} (\partial_i\zeta)^{2}\partial^2\zeta$. Using the in-in formula \eqref{Winin} \cite{Weinberg:2005vy},
\begin{align}
\langle \zeta_{\bfk_1}  \zeta_{\bfk_2} \zeta_{\bfk_3}\rangle
&  \supseteq
(2\pi)^3 \delta(\bfk_1+\bfk_2+\bfk_3) (\bfk_1\cdot\bfk_2) k_3^2\int_{-\infty}^0 \D \tau  \, \text{Im}\left[
\zeta_{k_1}(\tau) \zeta_{k_2}(\tau) \zeta_{k_3}(\tau) \zeta^*_{k_1}(0)\zeta^*_{k_2}(0)\zeta^*_{k_3}(0)
\right] 
\notag \\
&\qquad +\text{perms.}
\label{Winin-ex-app}
\end{align}
The integral is computed from the initial time (with standard $i\epsilon$-prescription switching off the interactions) up to late times. However, since we are interested in estimating the leading non-Gaussianities induced by the growing mode in the unstable phase, we can  use the fact that the integral is mostly dominated by the contributions coming from the interval $[\tau_0,0]$. Indeed, it is precisely the integral from $\tau_0$ to $0$ that involves the growing modes that are responsible for the exponential  enhancement of the spectrum and a pole in the flattened momentum configurations. Instead, the integral over $(-\infty,\tau_0]$ involves only products of  the mode functions \eqref{app:mode1}, yielding the standard total energy pole, without any exponential or polynomial enhancement in $\tau_0$ \cite{Arkani-Hamed:2015bza,Pajer:2020wxk}.  Thus, for our purposes, we can restrict the integral \eqref{Winin-ex-app} to the interval $[\tau_0,0]$,
\begin{align}
\langle \zeta_{\bfk_1}  \zeta_{\bfk_2} \zeta_{\bfk_3}\rangle
&  \supseteq
(2\pi)^3 \delta(\bfk_1+\bfk_2+\bfk_3) (\bfk_1\cdot\bfk_2) k_3^2\int_{\tau_0}^0 \D \tau  \, \text{Im}\left[
\zeta_{k_1}(\tau) \zeta_{k_2}(\tau) \zeta_{k_3}(\tau) \zeta^*_{k_1}(0)\zeta^*_{k_2}(0)\zeta^*_{k_3}(0)
\right] 
\notag \\
&\qquad +\text{perms.}
\label{Winin-ex-app-2}
\end{align}
The calculation then mimics the one in cosmologies with non-Bunch--Davies vacuum where both  positive and negative frequencies appear in the mode functions,
\begin{equation}
v_k(\tau) \simeq  \frac{   \e^{  - i c_s k \tau_0- \frac{i\pi}{4}} }{2 \sqrt{c_s k}}  \left[
 \e^{c_s k (\tau-\tau_0) } \left(1-\frac{1}{c_s k \tau }\right)+ i \,  \e^{-c_s k (\tau-\tau_0) } \left(\frac{1}{c_s k \tau }+1\right)
 \right] \, ,
\quad
\tau>\tau_0 \, ,
\label{vschem}
\end{equation}
for $k\vert\tau_0\vert\gg1$, 
and the result parallels the one of \cite{Garcia-Saenz:2018vqf,Bjorkmo:2019qno,Fumagalli:2019noh}. 
First, one can notice that, the $\zeta$ modes in \eqref{Winin-ex-app-2}, which we have to compute the   imaginary part of, given the solution \eqref{vschem}, cannot be all growing, but there must be instead an odd number of insertions of the decaying component.  
Then, one can identify two types of leading contributions. The first one arises from inserting one decaying mode in the external legs in   \eqref{Winin-ex-app-2}, while the internal legs are all growing. In this case, the integral between $\tau_0$ and $0$ provides a standard total energy pole, $k_T=k_1+k_2+k_3$ \cite{Arkani-Hamed:2015bza}, with no polynomials in $\tau_0$. As a result, the correlator  grows exponentially  as $\sim\e^{4 c_s k \vert\tau_0\vert}$ \cite{Garcia-Saenz:2018vqf,Bjorkmo:2019qno}.
The second leading contribution arises when the decaying mode is in one of the internal legs, while the external legs are all growing. In this case, the integral does not result in a term proportional to $1/k_T$, but it yields instead a pole in the flattened configuration \cite{Garcia-Saenz:2018vqf}. In this limit, in addition to the exponential enhancement $\sim\e^{4 c_s k \vert\tau_0\vert}$, the bispectrum contains a quartic polynomial in $\tau_0$, 
\begin{equation}
 \int_{\tau_0}^0 \frac{\D \tau}{\tau}  \, \text{Im}\left[
\zeta_{k_1}(\tau) \zeta_{k_2}(\tau) \zeta_{k_3}(\tau) \zeta^*_{k_1}(0)\zeta^*_{k_2}(0)\zeta^*_{k_3}(0)
\right]  
\supseteq -\frac{5 \tau_0^4 \e^{-2 c_s (k_2+k_3) \tau_0 }}{256c_s^6k_2^2 k_3^2 (k_2+k_3)^2}
\, ,
\label{Winin-ex-app-3}
\end{equation}
where the external momenta have been chosen in such a way that $k_1=k_2+k_3$. The generalization to higher-point correlation functions follows from analogous  considerations to those in \cite{Bjorkmo:2019qno,Fumagalli:2019noh,Ferreira:2020qkf}. The argument can be understood as follows. Considering a generic tree-level diagram, each vertex corresponds to one insertion of a certain interaction Hamiltonian and, therefore, to the integral of a certain commutator in the in-in formula   \eqref{Winin}. As noted above and pointed out in \cite{Bjorkmo:2019qno,Fumagalli:2019noh,Ferreira:2020qkf}, given the mode function in the form \eqref{vschem},  to have a non-zero contribution the fields in the imaginary part resulting from each commutator cannot be all of the same type. Instead, there  must be an odd number of both growing and decaying modes. In particular, the leading contribution corresponds to the case in which a single decaying mode is involved in each vertex of the diagram. As before, there are two possibilities: either the decaying mode is attached to an external field, or it is attached  to one of the internal legs of the vertex. In the first case, there is an overall exponential enhancement and the time integrals simply turn into some function of the external momenta. In the second case, the correlators are always suppressed, except for particular flattened configurations of the momenta. These configurations are such that  the decaying mode exactly cancels the growing exponentials and the result of  the integral is now some finite polynomial in $\tau_0$ \cite{Garcia-Saenz:2018vqf,Bjorkmo:2019qno,Fumagalli:2019noh,Ferreira:2020qkf}. At tree level, the degree of the polynomial depends on the number $n$ of external fields and the number $N$ of vertices in the diagram. For interaction Hamiltonians of the form $H_I^{(m_j)}= \int \D^3\bfx \, a^3 (\tau \partial_i \zeta)^{m_j-1} \tau^2 \partial_j^2\zeta $  (which are of the type discussed in the main text), where $j=1,\ldots,N$, we find---see also Eq.~\eqref{scalenpf-2} above---
\begin{equation}
\frac{\langle \zeta^n \rangle}{\langle \zeta^2 \rangle^{n-1}} \supseteq c_s^{3n-3}  \sum_{N} \frac{ \tau_0^{2n + 2N -4}}{c_s^{2n-N+1}}
 \, ,
 \label{app:sclng}
\end{equation}
where we used the tree-level identity $n= \sum_{j=1}^N m_j - 2 N + 2$  and the fact that $\langle \zeta^2 \rangle\sim c_s^{-3}$. 
The scaling in $\tau_0$ in \eqref{app:sclng} is maximum when $N=n-2$, which yields 
\begin{equation}
\frac{\langle \zeta^n \rangle}{\langle \zeta^2 \rangle^{n-1}} \sim c_s^{2n-6}  \tau_0^{4n-8} 
 \, ,
\label{app:sclng-2}
\end{equation}
in agreement with the explicit calculation \eqref{Winin-ex-app-3} for $n=3$.
Note that the enhancement is larger than the one estimated, e.g.\ in  \cite{Garcia-Saenz:2018vqf,Fumagalli:2019noh}, where the EFT was truncated at the leading order in derivatives. 
Note also that, had we chosen the interaction Hamiltonian $H_I^{(3)}= \int \D^3 \bfx \, a^3 \dot{\zeta}^3$ to compute the three-point function
\begin{equation}
\langle \zeta_{\bfk_1}  \zeta_{\bfk_2} \zeta_{\bfk_3}\rangle \propto
(2\pi)^3 \delta(\bfk_1+\bfk_2+\bfk_3) \int_{-\infty}^0 \frac{\D \tau}{\tau}  \, \text{Im}\left[
\zeta'_{k_1}(\tau) \zeta'_{k_2}(\tau) \zeta'_{k_3}(\tau) \zeta^*_{k_1}(0)\zeta^*_{k_2}(0)\zeta^*_{k_3}(0)
\right] \, ,
\end{equation}
we would have found \cite{Garcia-Saenz:2018vqf,Fumagalli:2019noh,Bjorkmo:2019qno,Fumagalli:2020adf,Ferreira:2020qkf}
\begin{equation}
 \int_{\tau_0}^0 \frac{\D \tau}{\tau}  \, \text{Im}\left[
\zeta'_{k_1}(\tau) \zeta'_{k_2}(\tau) \zeta'_{k_3}(\tau) \zeta^*_{k_1}(0)\zeta^*_{k_2}(0)\zeta^*_{k_3}(0)
\right]  
\supseteq \frac{ \tau_0^3 \e^{-2 c_s (k_2+k_3) \tau_0 }}{64c_s^3k_2 k_3 (k_2+k_3)}
\, ,
\end{equation}
which, as anticipated, is subleading compared with \eqref{Winin-ex-app-3}.

We conclude mentioning that the estimate \eqref{app:sclng-2} might be excessively pessimistic. In more realistic situations with smoother transitions between the various phases, there might be cancellations that result in a smaller enhancement of the correlators, see e.g.\ \cite{Taoso:2021uvl}.  However, these cancellations are model dependent and cannot be captured in full generality for arbitrary correlators.

\section{Large power spectra in multi-field models of inflation}
\label{app:2-fields}

In the main text, we discussed different scenarios in the context of the EFT of single-field inflation that can lead to an enhancement of the  final power spectrum for a range of momenta. As opposed to several previous works, we discussed here the role of higher derivative operators,   as well as of  approximate symmetries that ensure the robustness of the evolution.
In this appendix,   we review some known results in the context of multi-field models of inflation, which can be useful to contrast against the analysis that we have carried in this paper. We anticipate  that, even if it is possible to reduce these multi-field models to an effective single-field description at low energies, there are still important conceptual differences to keep in mind. For instance, it is not clear which sort of UV completion of multifield  models would give rise to the higher derivative operators, considered above, with WBG symmetry. This makes our analysis and the multi-field scenarios, e.g.\ of \cite{Garcia-Saenz:2018ifx,Garcia-Saenz:2018vqf, Fumagalli:2019noh, Bjorkmo:2019qno, Fumagalli:2020adf,Ferreira:2020qkf,Palma:2020ejf,Fumagalli:2020nvq}, substantially different. In particular,  in the scenario with negative $c_s^2$, the enhancement in the final power spectrum cannot in our discussion above be directly  related to the mass of some UV degree of freedom, as in the cases reviewed below, but it is instead  crucially related to the  scale $\Lambda_3$, at which the WBG operators enter.

\subsection{Two-field example}

{Let us consider a two-field model, with  $\phi^a$ denoting the field space coordinates}, whose components are labelled  by the index $a$. Let $V(\phi^a)$ be the potential, which depends on the  fields $\phi^a$. The kinetic term is $\gamma_{ab}\partial_\mu\phi^a\partial^\mu\phi^b$, where $\gamma_{ab}$ is the  metric on field space. 
It is convenient to  introduce the unit vectors $T^a\equiv \dot\phi_0^a/\dot\phi_0$ and $N_a\equiv \sqrt{\det \gamma} \, \epsilon_{ab}T^b $, where $\dot\phi_0\equiv \sqrt{\dot\phi_{0a}\dot\phi_0^a}$ and where $\epsilon_{ab}$ is  the two dimensional Levi-Civita symbol---see, e.g.\ \cite{Achucarro:2010da,Cespedes:2012hu,Achucarro:2012sm,Achucarro:2012yr} for further details.  $T^a$ and $N^a$ represent the tangent and normal to the trajectory \cite{Gordon:2000hv,GrootNibbelink:2000vx,GrootNibbelink:2001qt}, and  satisfy $T_aT^a=N_aN^a=1$ and $N_aT^a=0$.
In addition, we define
$D_t T^a \equiv \dot{T}^a+\Gamma^a_{bc}\dot{\phi}_0^bT^c= \Omega N^a$, 
where

$\Omega\equiv -N^a \partial_a V/\dot\phi_0$.

Slow-roll parameters are defined in analogy with  the single-field case, e.g.\ $\varepsilon\equiv \dot\phi_0^2/(2 \Mpl^2  H^2)$, where now $\phi_0$ depends on the two fields.

Following \cite{Cespedes:2012hu,Achucarro:2012yr}, defining the comoving curvature perturbation $\zeta$ and the (heavy) isocurvature perturbation $\psi$ by projecting $\delta\phi^a$ using the vectors $T^a$ and $N^a$, one finds the following quadratic action:
\begin{equation}
S=\frac{\Mpl^2}{2}\int \D^4 x \, a ^3\left[2\varepsilon\dot\zeta^2-\frac{2\varepsilon}{a^2}(\bfnabla\zeta)^2+\dot\psi^2-\frac{1}{a^2}(\bfnabla\psi)^2-M^2\psi^2-4\Omega\frac{\dot{\phi}_0}{H  \Mpl}\dot\zeta \psi \right]\label{eq:2fieldaction}
\end{equation}
where   $M$  is the mass of the  heavy field $\psi$ which  depends on background quantities,  $M^2= -\Omega^2+ N^aN^b \nabla_a\nabla_bV+H^2 \varepsilon \mathbb{R}$, with $ \mathbb{R}$ the Ricci scalar associated with the field space metric  $\gamma_{ab}$. Note that this expression allows  $M^2$  to be negative if, for example, the curvature $\mathbb{R}$ is negative, or if the coupling $\Omega$ is large enough.
It will be convenient to  introduce the speed of sound,
  \begin{align}
c_s^{-2}\equiv 1+\frac{4\Omega^2}{M^2} \, ,
\end{align}
which can be used to re-express  the ratio   $\Omega^2/M^2$. Its relation to \eqref{cs} (the sound speed we have used in this work) will be clear in a moment.  To see how the coupling $\Omega$ affects the dynamics of both fields, let us write the equations of motion \cite{Cespedes:2012hu}
\begin{align}
\ddot\zeta+(3+2\varepsilon-2\eta)H\dot\zeta+\frac{k^2}{a^2} \zeta&=2\Omega \frac{H\Mpl}{\dot\phi_0}\left[\dot\psi+\left(3-\eta+\varepsilon+ \frac{\dot{\Omega}}{H\Omega} \right)H\psi\right] \, ,
\label{appE:eom1}
\\
\ddot\psi+3H\dot\psi+\frac{k^2}{a^2}\psi+M^2\psi&=-2\frac{\dot\phi_0}{H\Mpl}\Omega\dot \zeta \, ,
\label{appE:eom2}
\end{align}
where we introduced $\eta\equiv -\ddot{\phi}_0/(H\dot{\phi}_0)$ and where we have   transformed the fields into Fourier space. In order to study the dynamics of the coupled equations above, it is convenient to make some approximations. In particular, we will focus on the high-energy limit, $k\gg aH$, where the dynamics of the modes is  well approximated by the flat-space limit, $H\rightarrow0$.  In this regime,  it is possible to ignore the friction terms and one can find a solution by using a plane wave ansatz, $\e^{i\omega t}$, for  the mode functions. Ignoring slow-roll terms, the  frequencies are given by \cite{Achucarro:2010jv,Baumann:2011su},
\begin{equation}
\omega_\pm^2=\frac{M^2}{2c_s^2}+\frac{k^2}{a^2}\pm\frac{M^2}{2 c_s^2}\sqrt{1+\frac{4c_s^2 k^2(1-c_s^2)}{a^2M^2}} \, ,
\end{equation}
which are in general non-analytic in $k$. 
In particular,  we are interested in the case where one of the two degrees of freedom is very heavy, in such a way that it can be integrated out and we can find an effective theory for the light mode. When $k\gg aH$, there is a  hierarchy between $\omega_-$ and $\omega_+$ if
\begin{equation}
\frac{k^2}{a^2}\ll\frac{M^2}{c_s^2} \, ,
\label{luv}
\end{equation}
which implies $\omega_-^2\ll \omega_+^2$.
%
%
Expanding $\omega_\pm$ in powers of $k/a$, we get
\begin{align}
\omega_-^2&=c_s^2\frac{k^2}{a^2}+\frac{(1-c_s^2)^2}{M^2c_s^{-2}}\frac{k^4}{a^4}+\mathcal{O}\left(\frac{k^6}{a^6}\right) \, ,
\label{eq:omega_}
\\
\omega_+^2&= M^2c_s^{-2}
+ (2-c_s^2)\frac{k^2}{a^2}-\frac{(1-c_s^2)^2}{M^2c_s^{-2}}\frac{k^4}{a^4}
+\mathcal{O}\left(\frac{k^6}{a^6}\right) \, ,
\end{align}
where we neglected   terms  suppressed by higher powers of  $c_sk/(aM)$.
In the following, we are interested in the regime $M\ll \Omega$, which corresponds to $c_s^2\ll1$. Note that when $M^2 \lesssim \frac{k^2}{a^2}  \ll M^2c_s^{-2}$, the second term in \eqref{eq:omega_} dominates and the dispersion relation of the light mode becomes of the form $\omega^2_- \approx k^4/(4\Omega^2a^4)$,  resembling  the regime discussed in Section \ref{sec:srtogis}. 
We recover instead a linear dispersion relation with a small sound speed when $ \frac{k^2}{a^2} \ll M^2   \ll M^2c_s^{-2}$.
Following the nomenclature of  \cite{Baumann:2011su,Gwyn:2012mw}, one can define 
\begin{equation}
\Lambda_{\rm{new}}=M c_s \, ,
\end{equation}
which corresponds to the energy scale at which the dispersion relation for $\omega_-$ turns from linear to quadratic, to be compared with $\Lambda_\text{UV}\equiv M c_s^{-1}$, which is instead the scale where the single-field effective description is no longer applicable and one should integrate back in the heavy field.\footnote{$\Lambda_\text{UV}$ can be simply obtained by plugging $k_\text{UV}/a=M/c_s$, given by \eqref{luv}, into \eqref{eq:omega_}, when the $k^4$-term dominates.}

After having established the relevant scales, we shall go back to the quadratic action \eqref{eq:2fieldaction} and, under the assumption of \eqref{luv},  integrate out the massive mode explicitly to  obtain an effective description of the light degree of freedom.
Using the equation \eqref{appE:eom2}, in the regime where one neglects  time derivatives of $\psi$,
\begin{equation}
\vert\ddot\psi\vert\ll M^2\psi,\qquad H^2\ll M^2 \, ,
\label{stdpsi}
\end{equation}
we can solve for $\psi$ as a function of  the curvature mode as\footnote{In deriving this equation we assumed that time derivatives on $\psi$ are subdominant. This can be translated into a constraint over the coupling $\Omega$. Assuming  that the main time dependence in $\psi$, given in \eqref{eq:psirelation}, comes from the coupling  $\Omega$,  the condition \eqref{stdpsi} becomes $\vert\ddot{\Omega}\vert\ll M^2 \vert\Omega\vert$, where we neglected time derivatives of the other background quantities.
Now assuming that the typical  time scale of $\Omega$ is given by  $\Delta N/H$, where $N$ denotes the number of e-folds, one finds the condition $\Delta N\gg H/ M$, as it should be.}
\begin{equation}
\psi=-\frac{2\Omega\dot\phi_0}{H(k^2/a^2+M^2)}\dot\zeta \label{eq:psirelation}
\end{equation}
Plugging  back into the action we get,
\begin{equation}
S=\frac{\Mpl^2}{2}\int \D t \D^3\bfk  \,  a^3 2\varepsilon \left[\left(\frac{M^2 c_s^{-2}+k^2/a^2}{M^2+k^2/a^2}\right)\dot\zeta^2-\frac{k^2}{a^2}\zeta^2\right] \, .
\end{equation}

The equation of motion for $\zeta$ can be written in the form \cite{Baumann:2011su,Achucarro:2010da,Gwyn:2012mw},
\begin{equation}
\ddot \zeta+ 3H \dot\zeta+ \frac{2H (1-c_s^2)M^2 k^2/a^2}{(M^2+k^2/a^2)(M^2+c_s^2 k^2/a^2)}\dot\zeta
+ \frac{M^2+k^2/a^2}{M^2 c_s^{-2}+k^2/a^2} \frac{k^2}{a^2} \zeta=0 \, .
\label{eq:nonlocalcs}
\end{equation}
Notice that when $c_s=1$ the `non-local' terms disappear and one recovers the usual equation of motion for $\zeta$. However, in the general case, when $c_s\neq1$, the equation is rather complicated. To find an approximate analytic solution, we will work, in the following, in the regime  of small $c_s$. In the limit $c_s^2\ll 1$, Eq.~\eqref{eq:nonlocalcs} becomes
\begin{align}
\ddot \zeta+ \left(3H+\frac{2H k^2}{k^2+M^2 a^2}\right)\dot\zeta+\left(c_s ^2\frac{k^2}{a^2}+\frac{k^4}{a^4 M^2 c_s^{-2} }\right)\zeta=0 \, ,
\label{eq:nonlocalcs-2}
\end{align}
where we neglected subleading orders in powers of $c_s^2$.
Let us now look at \eqref{eq:nonlocalcs-2} in the two regimes considered above.
Let us consider modes with wavenumber $k$  that, at sufficiently early times, satisfy the condition $M^2\ll k^2/a^2\ll M^2 c_s^{-2}$. For these modes, the  $k^4$ term in \eqref{eq:nonlocalcs-2} dominates and the dynamics is effectively described by the equation
\begin{align}
\ddot \zeta+ 5H\dot\zeta+\frac{c_s^{2}k^4}{a^4 M^2  }\zeta=0,\qquad k/a\gg M \, .
\label{eee1}
\end{align}
As the system evolves in time, the $k^4$ contribution in \eqref{eq:nonlocalcs-2} redshifts faster than the other $k^2$ term, until it becomes subdominant. At that point, when  $k/a< M$, the equation of motion takes on the standard form
 \begin{align}
 \ddot \zeta+ 3H\dot\zeta+c_s^2 \frac{ k^2}{a^2}\zeta=0,\qquad k/a\ll M \, .
 \label{eee2}
\end{align}
Thus, instead of solving  \eqref{eq:nonlocalcs} exactly, we can study these two asymptotic regimes and solve \eqref{eee1} and \eqref{eee2} separately, and   then perform the matching  at the crossing point $a=k/M$.

Defining the canonically normalized fields  $u = \frac{\sqrt{2\varepsilon}\,\Mpl M}{c_s k} a^2\zeta $ and $v = \frac{\sqrt{2\varepsilon}\Mpl}{c_s}a\zeta$ in the two regimes, the equations \eqref{eee1} and \eqref{eee2} read
\begin{align}
u_k''+\left(\frac{c_s^2  H^2}{M^2}k^4\tau^2-\frac{6}{\tau^2}\right)u_k=0 \, ,\qquad k/a\gg M \, ,
 \label{eq:speeda}
 \\
v_k''+\left(c_s^2k ^2-\frac{2}{\tau^2}\right)v_k=0 \, , \qquad k/a\ll M \, .
\label{eq:speedb}
\end{align}
To solve \eqref{eq:speeda}, which describes the short-distance dynamics of the perturbations, we should impose the correct Bunch--Davies condition. Selecting the positive frequency solution in the limit $k/a\gg M $, one finds \cite{Baumann:2011su}
\begin{equation}
 u_{k}(\tau)=\sqrt{\frac{\pi}{8}}(-\tau)^{1/2}H^{(1)}_{5/4}\left(\frac{Hc_s}{2M}(k \tau)^2\right) \, .
\label{eq:modefunctionLowcs}
 \end{equation} 
At late times, when the $k^2$ term dominates in the dispersion relation, the solution to \eqref{eq:speedb} for $v_k$ can be written in general as
\begin{equation}
v_k(\tau)=\sqrt{-\tau}\left(A_k H^{(1)}_{3/2}(c_s k\tau)+B_k H^{(2)}_{3/2}(c_s k\tau)\right) \, ,
\label{vkmfs}
\end{equation}
where the coefficients $A_k$ and $B_k$ can be fixed by requiring that $\zeta$ and $\zeta'$ are continuous across the transition point $\tau=-M/(H k)$.
Note that in the case of \eqref{eq:speeda}, Hubble crossing occurs at $k^2\tau_c^2 \sim M/(Hc_s)$, while for \eqref{eq:speedb} it happens at $\vert k\tau_c\vert \sim 1/c_s$. Let us start considering the case in which Hubble crossing happens   in the phase where the dynamics is described by \eqref{eq:speedb}, i.e.\ $H/(M c_s)\ll 1$.
After matching \eqref{vkmfs} with \eqref{eq:modefunctionLowcs} at $\tau=-M/(H k)$, one finds the following 
power spectrum,
\begin{equation}
\Delta_\zeta^2 = \frac{k^3}{2\pi^2} \langle \zeta_k\zeta_{-k}\rangle' = \frac{(H/\MP)^2}{8 \pi ^2  \varepsilon c_s} \, .
\label{eq:2ptTra}
\end{equation}

Note that the power spectrum takes the usual form, where the dependence on $M$ is only through $c_s$, as it should be.

In the opposite regime instead,  when $M c_s \ll H$, modes exit the horizon in the phase described by \eqref{eq:speeda}, where the quartic term  dominates. Thus, using \eqref{eq:modefunctionLowcs}, the  
power spectrum is found to be \cite{Gwyn:2012mw}

\begin{equation}
\Delta_\zeta^2  = \sqrt{\frac{M}{c_s H}}\Gamma(5/4)^2\frac{(H/\MP)^2}{\pi^3\varepsilon} \, .
\label{eq:appgi}
\end{equation}

Note that, in this case, the power spectrum is enhanced by a factor of $\sim\sqrt{M c_s/H}=c_s\sqrt{\Lambda_\text{UV}/H}$, with respect to \eqref{eq:2ptTra}.
In this regime, the strong coupling scale is found to be \cite{Gwyn:2012mw}, 
\begin{equation}
\Lambda_\star = \left(\frac{2}{\pi}\right)^{2/5} \left(\frac{ H^2}{8\pi^2 \MP^2 \varepsilon} \right)^{-2/5} \left(\frac{H^4c_s^2}{\Lambda_\text{UV}^4}\right)^{2/5} \Lambda_\text{UV} \, ,
\label{eq:mfsUV}
\end{equation}

Now, similarly to the case discussed in Section~\ref{sec:srtogis}, one can imagine a situation where the two-field model  can be initially described at low energies in terms of an effective theory for a single degree of freedom with $c_s\sim1$. Then, the UV dynamics is such that $c_s$ evolves in time and becomes small. In particular, we shall assume that $c_s$ becomes small enough that some of the modes end up being governed by a quadratic dispersion relation.
In this scenario, there will be modes with sufficiently small $k$ that are still in the initial phase with $c_s\sim1$ when they  exit the horizon. These will have a final power spectrum given by \eqref{eq:2ptTra} with $c_s\sim1$. On the other hand,  modes with larger $k$ that cross the horizon in the phase dominated by \eqref{eq:modefunctionLowcs}\footnote{To be precise, in the scenario, one should reconsider the solution \eqref{eq:modefunctionLowcs} and the Bunch--Davies early-time condition. However, if the crossover between the phases with $c_s\sim1$ and $c_s\ll 1$ happens when $\vert k\tau\vert \gg1$, this will affect the 
power spectrum of such modes    by an irrelevant order-one factor---see also Section~\ref{sec:srtogis}.} will have a final power spectrum enhanced, compared to the former modes, by the factor $\sqrt{\Lambda_\text{UV}/H}$. Clearly, this factor cannot be arbitrarily large---or, $c_s$ cannot be arbitrarily small---as it should satisfied the condition $\Lambda_\star\gtrsim\Lambda_\text{UV}$, with $\Lambda_\star$ given in \eqref{eq:mfsUV}. However, already for $c_s\sim10^{-1}$, one finds  $\sqrt{\Lambda_\text{UV}/H}\lesssim 10$, which provides a strong limitation on the possible enhancement of the power spectrum.

\subsection{Transient phase with imaginary  sound speed}

As discussed e.g.\ in \cite{Garcia-Saenz:2018ifx,Garcia-Saenz:2018vqf, Fumagalli:2019noh, Bjorkmo:2019qno, Fumagalli:2020adf,Ferreira:2020qkf},

one can obtain a significantly larger power spectrum if the heavy field turns out to be tachyonic, $M^2<0$, which, in the low-energy description corresponds to having a gradient instability, $c_s^2<0$, (we are still assuming $\vert c_s^2\vert \ll1$) in the dynamics of the light mode.

Replacing $c_s^2\mapsto-c_s^2$ and $M^2\mapsto-M^2$ in \eqref{eq:nonlocalcs-2}, we can write the mode function equation as
\begin{align}
\ddot \zeta+ \left(3H+\frac{2H k^2}{k^2-M^2 a^2}\right)\dot\zeta+\left(-c_s ^2\frac{k^2}{a^2}+\frac{k^4}{a^4 M^2 c_s^{-2} }\right)\zeta=0 \, .
\end{align}
The effect of the instability becomes visible only when $k/a \ll \vert M\vert$.
In this limit, the dynamics is effectively described by
\begin{align}
 v''_k+ \left(-c_s^2 k -\frac{2}{\tau^2}\right)v_k=0 \, ,
\end{align}
which admits the general solution 
\begin{align}
v_k=(-\tau)^{1/2}\left(A_k H^{(1)}_{3/2}(-i c_s k\tau)+B_k H_{3/2}^{(2)}(-i c_s k \tau)\right) \, .
\label{eq:modefunctiontac}
\end{align}
For larger $k$, more precisely when $k/a \gg \vert M\vert$,  the theory is instead dominated by the $k^4$ term and the dynamics is stable.  The equation of motion is  the same as \eqref{eq:speeda}, with the solution \eqref{eq:modefunctionLowcs}.

Following the same procedure we described above, the 
power spectrum is found to be
\begin{equation}
\Delta_\zeta^2 = \frac{k^3}{2\pi^2} \langle \zeta_k\zeta_{-k}\rangle' = \frac{(H/\MP)^2}{16 \pi ^2  \varepsilon c_s} \, \e^{\frac{2c_s M}{H}} \, ,
\label{eq:2ptTracs0}
\end{equation}
where the quantity in the exponent has to be $\vert c_s M/H\vert \gg1 $ in order for  Hubble crossing to happen in the unstable phase. Otherwise, the modes would not be significantly  enhanced by the instability and the power spectrum would still be given by \eqref{eq:appgi}.\footnote{In \cite{Bjorkmo:2019qno}, using a WKB approach, 
the exponential enhancement in the power spectrum is estimated as $\e^{2x}$, where $x$ is found  to be
$ x=\frac{\pi}{2}\left(2-\sqrt{3+\xi}\right)\frac{\Omega}{H}$,
 where $\xi=\frac{1+4c_s^2}{1-c_s^2}$. Expanding this expression for small $c_s$, one finds that $x=\frac{\pi}{2} \frac{M c_s}{H}$, which agrees, up to an order-one factor, with \eqref{eq:2ptTracs0}.}

\bibliographystyle{hunsrt}
\bibliography{biblio_v16_GB}

\end{document}